\documentclass[10pt]{article}
\usepackage{amssymb,amsmath}
\usepackage{graphicx}
\usepackage{color}
\usepackage{hyperref}
\usepackage{curves,multicol}
\usepackage{geometry}
\usepackage{times}
\usepackage{bm}
\usepackage{ulem}
\usepackage{ucs}
\usepackage[utf8x]{inputenc}
\usepackage[intoc]{nomencl}
\usepackage{float}
\usepackage{dcolumn}
\usepackage{units}
\usepackage{titlesec}
\usepackage{curves,multicol}
\usepackage{mathptmx}
\usepackage{lscape}
\usepackage{nopageno}
\usepackage[labelfont=bf]{caption}
\usepackage{lipsum}   
\usepackage[none]{hyphenat} 

\makeatletter

\renewcommand\@biblabel[1]{}

\renewcommand{\section}{\@startsection
{section}
{1}
{0mm}
{-\baselineskip}
{0.5\baselineskip}
{\normalfont\bfseries\MakeUppercase}} 

{

\renewcommand{\subsection}{\@startsection
{subsection}
{2}
{0mm}
{0.5\baselineskip}
{0.25\baselineskip}
{\bfseries\normalsize}} 

\makeatother

\begin{document}
\sloppy

\setcounter{secnumdepth}{-1} 

\vspace*{-1.0cm}
\begin{flushright} \vbox{
32$^{\mathrm{nd}}$ Symposium on Naval Hydrodynamics\\
Hamburg, Germany, 5-10 August 2018}
\end{flushright}

\vskip0.65cm
\begin{center}
\textbf{\LARGE
High Fidelity Simulations of Micro-Bubble Shedding from Retracting Thin Gas Films in the Context of Liquid-Liquid Impact\\[0.35cm]
}

\Large S. Mirjalili, W.H.R. Chan, A. Mani\\ 

(Department of Mechanical Engineering, Stanford University, Stanford, CA 94305, USA)\\
\vspace*{0.25cm}

\end{center}

\begin{multicols*}{2}

\section{Abstract}
\label{SECabstract}
Micro-bubbles are of significant interest due to the long-living signature they leave behind naval ships. In order to numerically model and predict these bubbles in naval applications, subgrid-scale models are required because of the extreme separation of length- and time-scales between the macroscopic geometries and the physical processes leading to the formation of these bubbles. Yet, there is much that is unknown about the mechanism behind the entrainment of such bubbles. Furthermore, quantitative information regarding their size distribution and dependence on flow parameters is very limited. Impact events are hypothesized to be the main contributor to the generation of micro-bubbles. This is due to a phenomenon known as Mesler entrainment, which has been observed in the context of the drop-pool impact problem. Namely, when a water droplet with diameter of $\mathcal{O}(1mm)$ impacts a deep water pool with an impact velocity of $\mathcal{O}(1m/s)$, hundreds of air micro-bubbles are entrained into the pool. These bubbles have been found to be remnants of a very thin air film entrapped between the two liquid bodies. These films have extremely high aspect ratios and after being punctured, shed micro-bubbles while retracting on time scales much shorter than the outer flow time scales. 

This separation of time-scales, along with a lacking of studies on retracting thin gas films has motivated us to study this fundamental problem with numerical simulations in two and three dimensions. Using a diffuse interface method, we perform two-phase simulations of retracting thin gas films in initially static liquid backgrounds to gain understanding regarding this problem and gather statistics that can be potentially used in a subgrid-scale model to predict micro-bubble shedding from a thin gas film. Coupling a model like this to a model that predicts the thin film characteristics resulting from an impact event, and a real-time impact detection algorithm, allows for predicting micro-bubbles within large scale two-phase flow simulations of naval interest.
\section{Introduction}
\label{SECintroduction}
A wide range of bubbles can be found in the wake of naval ships (Trevorrow \textit{et al.} 1994, Reed \& Milgram, 2002). Among these bubbles, bubbles in the range of $10\mu m-100\mu m$ are of particular interest as they neither dissolve nor rise to the surface (via buoyancy) fast enough to disappear from the trail of ships. In fact these bubbles, which we denote as micro-bubbles, remain under water for hours, leading to an easily detectable trail behind ships. 

\begin{figure}[H]
\centering
\includegraphics[width=\linewidth]{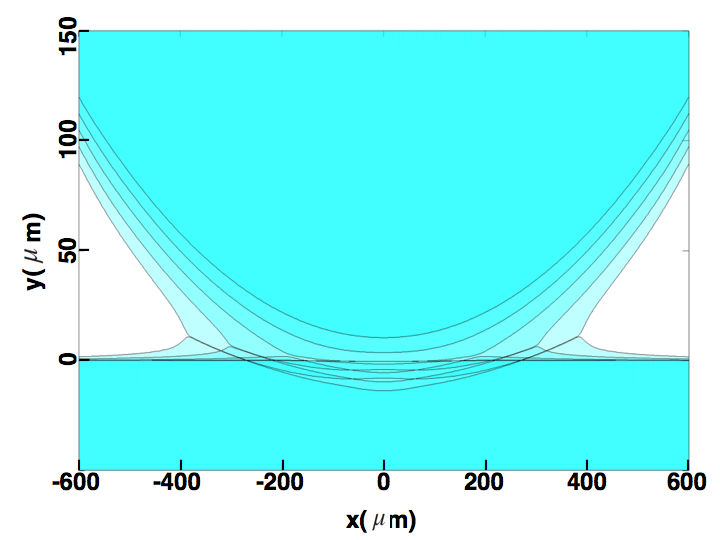}
\caption{Evolution of the water-air interfaces for a drop impacting a deep pool obtained from boundary integral method. Drop diameter is $D=3.4mm$ and impact velocity is $U=1m/s$.} 
\label{timeevltn}
\end{figure}

Within literature, the Mesler entrainment phenomenon in the context of the drop-pool impact problem is the best known instance where air micro-bubbles can be observed (Esmailzadeh \& Mesler 1986, Thoroddsen \textit{et al.} 2003, Saylor \& Bounds 2012, Thoroddsen \textit{et al.} 2012). Mesler entrainment happens when a water drop with diameter of $\mathcal{O}(1mm)$ impacts a deep water pool with an impact velocity of $\mathcal{O}(1m/s)$, to leave hundreds of air micro-bubbles in the pool. For such drop-pool impact events, there are multiple stages and different possible outcomes. First, as the air mediating the drop and pool is squeezed out and the gas layer becomes thin enough to around $\mathcal{O}(1\mu m)$, a high pressure zone builds up in the center of the film, "pumping" air out of the thin gas film. It is well known from experiments (Thoroddsen \textit{et al.} 2003, Tran \textit{et al.} 2013) and numerical simulations (Hicks \& Purvis 2011, Hendrix \textit{et al.} 2016) that this high pressure deforms the drop and pool, leading to two "kinks" that move radially away from the center. In Figure \ref{timeevltn} the formation of the kinks and deformation of the two liquid surfaces can be seen from a numerical simulation using boundary integral method (Mirjalili \& Mani, 2014). Experimental evidence from Thoroddsen \textit{et al.} (2012) shows that Mesler entrainment happens when the thin gas film prevents coalescence of the drop and pool while spreading around the drop such that a very thin hemispherical cap with thickness of the order of $\mathcal{O}(100nm)$ separates the two liquid bodies. At this stage, the very thin gas film ruptures due to intermolecular forces. The thin air film then retracts from the rupture points due to capillary forces and sheds micro-bubbles along the way. Some of the remnant air also form micro-bubbles when films from different rupture points come together to form threads of micro-bubbles. These steps can be clearly seen from experimental images of Thoroddsen \textit{et al.} (2012) .      
Naval ships can induce bow waves, transom waves and waves due to turbulent boundary layers along the ship hull. Such waves often become turbulent breaking waves, resulting in entrainment of air in complicated processes occurring at a wide range of time scales and length scales. Recent numerical studies of breaking waves have been successful at resolving scales down to below the Hinze scale (Mortazavi \textit{et al.} 2016, Wang \textit{et al.} 2016, Deike \textit{et al.} 2016), enhancing our understanding of such flow structures. However, the smallest micro-bubbles cannot be captured via such simulations as they often involve scales multiple orders of magnitude below the Hinze scale. The dynamics of breaking waves take place on $\mathcal{O}(1m)$ and $\mathcal{O}(1s)$ length and time scales respectively. The resolved Hinze scale, indicating the scale at which turbulent forces are in competition with surface tension forces,  is on the order of $\mathcal{O}(1mm)$, with time scale $\mathcal{O}(1ms)$. On the other hand, the events during liquid-liquid impacts and the subsequent Mesler entrainment at least require resolving events happening at length scales of $\mathcal{O}(100nm)$ and time scales of $\mathcal{O}(100ns)$. As such, subgrid-scale (SGS) models such as the proposed method by Chan \textit{et al.} (2016) and Chan \textit{et al.} (2017) are necessary to capture and predict micro-bubbles in turbulent breaking waves simulations. Based on the assumption that Mesler entrainment is the mechanism responsible for micro-bubble generation, this type of SGS model uses the drop-pool impact problem as a model problem to predict the micro-bubbles in the flow. Particularly, such SGS algorithms would map any collision between two liquid bodies to a drop-pool impact event via relevant non-dimensional numbers. For a water-air system, the non-dimensional parameters that can describe a drop-pool impact event are the Weber ($We={\rho}_{l}{U}^{2}D/\sigma$) number and Stokes ($St={\rho}_{l}UD/{\mu}_{g}$) number, where subscripts of "l" and "g" denote properties of the liquid and gas, respectively, $U$ is the impact velocity, $D$ is the diameter of the drop, $\sigma$ is the surface tension, $\rho$ is density and $\mu$ is viscosity. 



\begin{figure}[H]
\centering
\includegraphics[width=\linewidth]{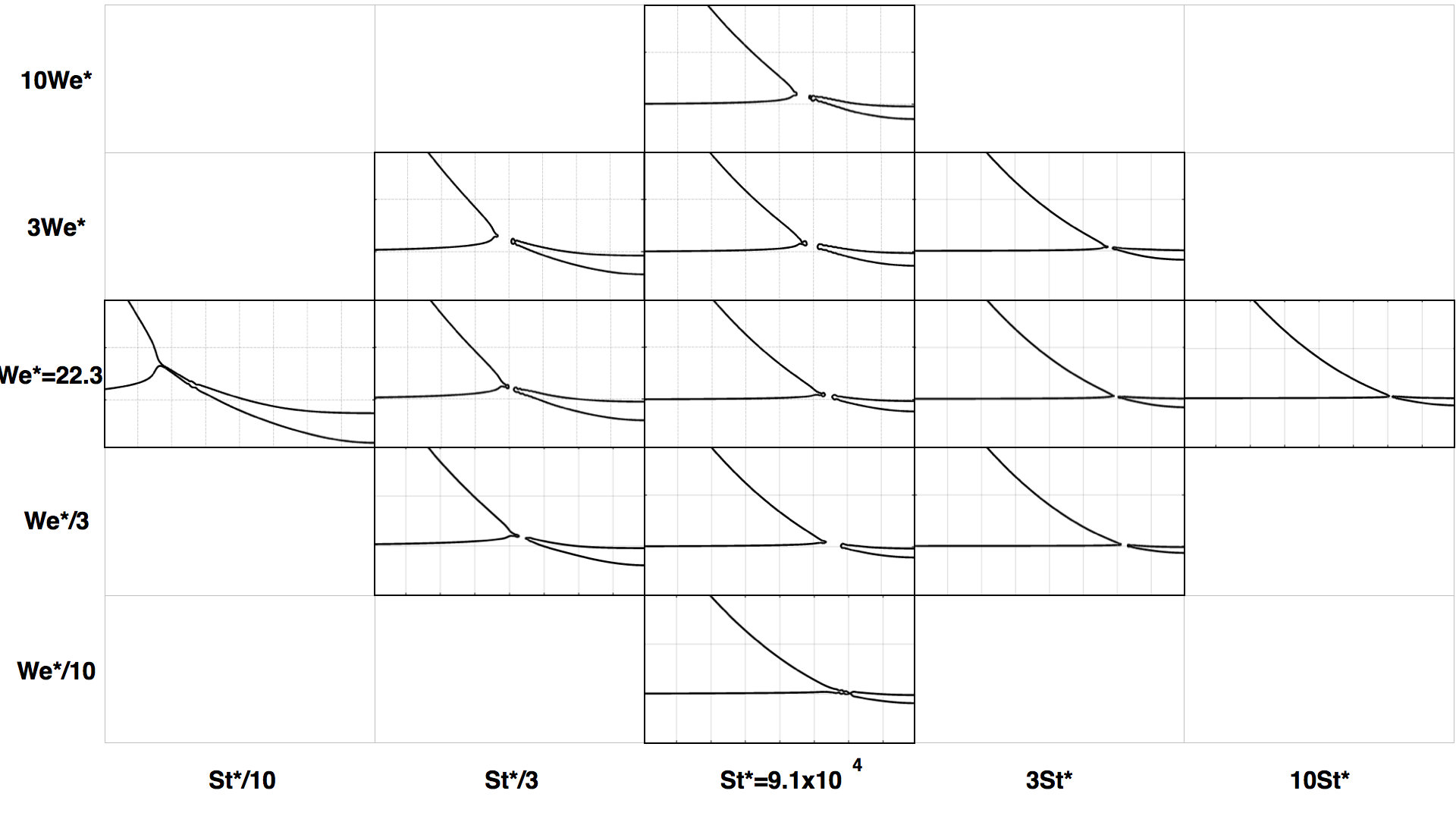}
\caption{Thin entrapped film after contact for a drop impacting a deep pool in an air-water system is plotted as a function of $St$ and $We$ of impact. In each plot the right boundary represents the axis of symmetry. Results obtained from high resolution diffuse interface calculations introduced in Mirjalili \textit{et al.} (2018b). } \label{table_self_similar}
\end{figure}

As mentioned before, numerical and experimental efforts focused on capturing the thin gas film entrapped during drop-pool impact are available in the literature. We have also investigated this problem via a diffuse interface interface-capturing scheme introduced in Mirjalili \textit{et al.} (2018a). Figure \ref{table_self_similar} shows the outcome of such simulations across a wide range of $St$ and $We$ numbers relevant to the Mesler entrainment regime. From these simulations and experimental evidence, the film thickness in drop-pool impact events in the Mesler entrainment regime is from $\sim 100nm$ to $\sim 1\mu m$. With that, the only step of the process that is still not well-understood is the shedding or generation of micro-bubbles when these films retract. Hence, in this work we will focus on the dynamics of a retracting gas film in a background liquid. 
After rupture of the thin gas film, a very high surface tension force field at the rim of the film causes film retraction on time-scales much smaller than the background flow time scale. It is thus justified to disregard the effects of the background flow. Moreover, since the film thickness is multiple orders of magnitude smaller than the radius of the drop, which is of the order of the film length, the curvedness of the film can also be neglected. Thus, we can effectively study the post-rupture dynamics of the hemispherical thin gas film entrapped during drop-pool impact events via analyzing a planar gas film retracting in an initially quiescent liquid. It is important to point out that the reverse of this problem, the retracting liquid film, is a well-studied classical problem that has received attention due to its role in formation of fingers and secondary drops in impact problems and atomization (Brenner \& Gueyffier 1999, S\"{u}nderhauf \textit{et al.} 2002, Roisman \textit{et al.} 2006, Savva \& Bush 2009, Krechetnikov 2010, Agbaglah \textit{et al.} 2013). Nevertheless, there is little to no quantitative understanding on the problem of a retracting gas film in a liquid. These two problems are significantly different. In the case of retracting liquid films, the effects of the background gas phase on the momentum transport can be and has been ignored in most analyses. For retracting gas films, however, the high-inertia flow in the background liquid drives the evolution of the film. In the present study we numerically study this problem. These simulations help us gain understanding of the relevant instabilities and the mechanism leading to micro-bubble shedding.


In the following, we will first formulate the problem and describe the numerical approach we have taken in this work for studying retracting thin gas films in liquid surroundings. Next, we will analyze 2D retracting gas films with aid from numerical simulations. After establishing the necessity for considering 3D effects, we will present our analysis and numerical results from high-fidelity 3D simulations of retracting gas films. Finally, we conclude this work with a summary of our findings.

\section{Problem Formulation and Numerical Approach}
\label{SECpfna}
\begin{figure}[H]
\centering
\includegraphics[width=\linewidth]{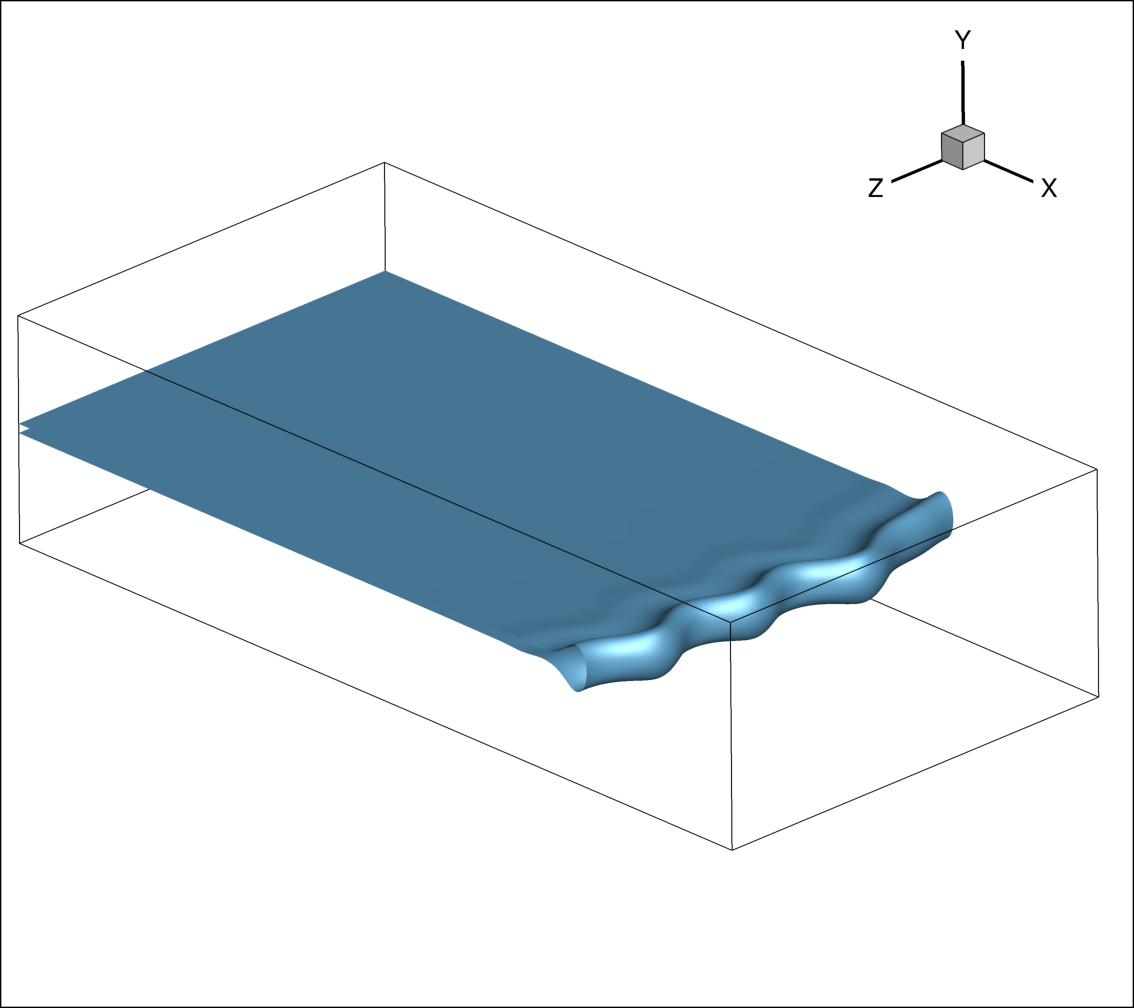}
\caption{A sample snapshot of thin gas film simulations showing the axis and typical domain size.} \label{sample_image}
\end{figure}

As explained above, we disregard the curvature of the film and the effect of background flow in our study of retracting thin gas films. Thus, we examine a planar retracting thin gas film with thickness $h$ and length $L$ in an initially stationary liquid surrounding. From experiments and numerical results, the thin gas films have very high aspect ratios such that $L\gg h$. The dynamics of a long thin air film in a background of water is determined only by its thickness. Thus, there is only one non-dimensional number that controls this problem for a given pair of working fluids (i.e. density and viscosity ratios). In this work, similar to literature on retracting liquid films, we have chosen to use the Ohnesorge number ($Oh={\mu}_{l}/\sqrt{{\rho}_{l}h\sigma}$) as the key dimensionless parameter to differentiate between different settings. In all our simulations we ensure that the domain and film length are large enough to correctly capture the evolution of the edge (rim) of the retracting gas film. In all simulations, the film is aligned with the x axis and retracts towards the negative x direction, the y axis is normal to the top and bottom sides of the film and the z direction is in the span-wise direction, as shown by a simulation snapshot in Figure \ref{sample_image}. This figure also shows a typical domain size used for our simulations. We will explain initial conditions, boundary conditions and mesh requirements for each simulation as we present the results in the following sections.

We examine the evolution of retracting thin gas films through incompressible two-phase flow simulations using a phase field method introduced in Mirjalili \textit{et al.} (2018a). The conservative phase field equation given by 
 \begin{equation}
\frac { \partial \phi  }{ \partial t } +\nabla \cdot \left(\vec{ u } \phi \right)=\nabla \cdot \left[\gamma \left(\epsilon \nabla \phi -\phi \left(1-\phi \right)\frac { { \nabla \phi  }  }{ \left| { \nabla \phi  }  \right|  }  \right) \right],
\label{phitrans}
\end{equation}
is discretized using central differences on staggered grids where $\phi$ represents the volume fraction of liquid in each phase such that $\phi=1$ in pure water and $\phi=0$ in pure air. $\epsilon$ and $\gamma$ are the free parameters of the numerical setup, respectively representing the interface thickness and interface formation speed. Realistic air-water interfaces are sharp and are formed at speeds much faster than the outer hydrodynamic scale; directly capturing their scale would lead to enormously stiff numerical systems. An alternative asymptotic approach is to select finite $\epsilon$ and $\gamma$ such that the numerical interfacial thickness, while resolvable, remains much smaller than any hydrodynamic geometrical scale, and its formation speed remains much faster than the flow scale. Mirjalili \textit{et al.} (2018a) developed a detailed investigation of the $\epsilon$-$\gamma$ parameter space and identified zones leading to robust, bounded, and well-posed numerical models for interfaces. Following their recommendation, we selected a fine mesh with $\epsilon/\Delta x=\gamma/{|\vec{u}|}_{max}=1$. This choice guarantees the boundedness of the phase field and obviates clipping or mass redistribution, as explained in Mirjalili \textit{et al.} (2018a). Continuity for incompressible flow is expressed by 
\begin{equation}
\nabla\cdot\vec{u}=0,
\label{continuity}
\end{equation}
and the Navier-Stokes equations can be written as
\begin{equation}
\frac{\partial \vec{u}}{\partial t}+\nabla\cdot(\vec{u}\otimes\vec{u})=\frac{1}{\rho}\left(-\nabla P+\nabla\cdot[\mu(\nabla \vec{u}+\nabla^{T}\vec{u})]+\sigma\kappa\nabla\phi\right)
\label{NS},
\end{equation}
where curvature ($\kappa$) is computed via
\begin{equation}
\kappa=\nabla\cdot\vec{n},
\label{curvature}
\end{equation}
and both density and viscosity are linear with respect to $\phi$,
\begin{equation}
\rho=(\rho_{l}-\rho_{g})\phi+\rho_{g},
\label{rho}
\end{equation}
\begin{equation}
\mu=(\mu_{l}-\mu_{g})\phi+\mu_{g}.
\label{mu}
\end{equation}
Equations \ref{NS} to \ref{mu} are time-integrated along with Equation \ref{phitrans} using a second order fractional step method requiring only one Poisson solve per time-step (Le \& Moin, 1991). The Poisson solve is performed using an efficient state of the art in-house parallel multigrid solver (Esmaily \textit{et al.}, 2017). The accuracy, efficiency and robustness of this fully-coupled two-phase solver was studied in Mirjalili \textit{et al.} (2018b).
\section{2D Retracting thin gas films}
\label{SEC2D}
2D simulations are much cheaper than their 3D counterparts, but offer valuable information and important understandings of retracting thin gas films. We first use 2D simulations to compare the dynamics of retracting gas films with liquid films. Then, looking at the retraction of long films we explore the possibility of micro-bubble shedding from 2D films and their retraction speed. By varying the $Oh$ number, we examine how the film thickness affects the dynamics for a water/air system. It is noteworthy that by mesh refinement in 2D, we also extract the resolution requirements for our 3D studies.

\begin{figure}[H]
\centering
\includegraphics[width=\linewidth]{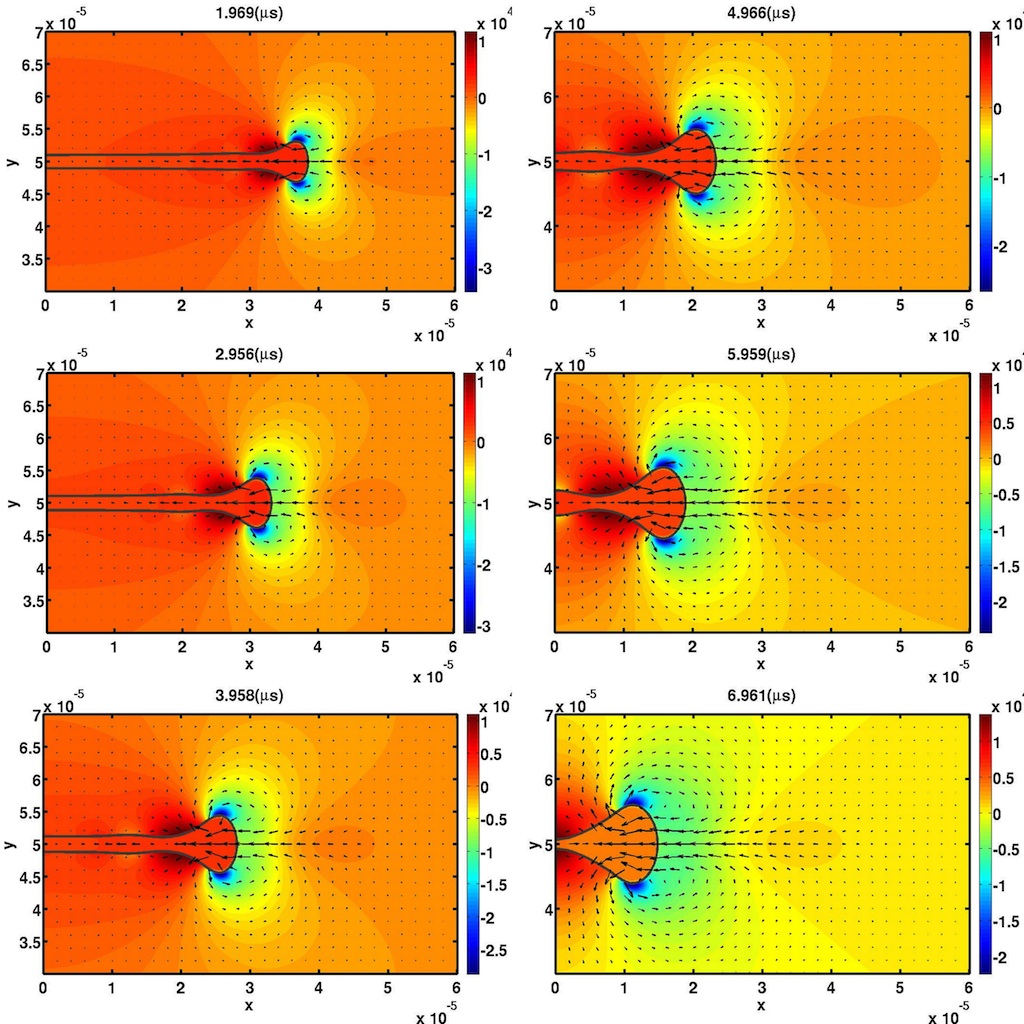}
\caption{Thin air film retracting in water. Pressure is shown in color and velocity field is drawn on top. Counter rotating vortices drive the film to the left of the domain.} 
\label{2D_gas_film}
\end{figure}

\begin{figure}[H]
\centering
\includegraphics[width=\linewidth]{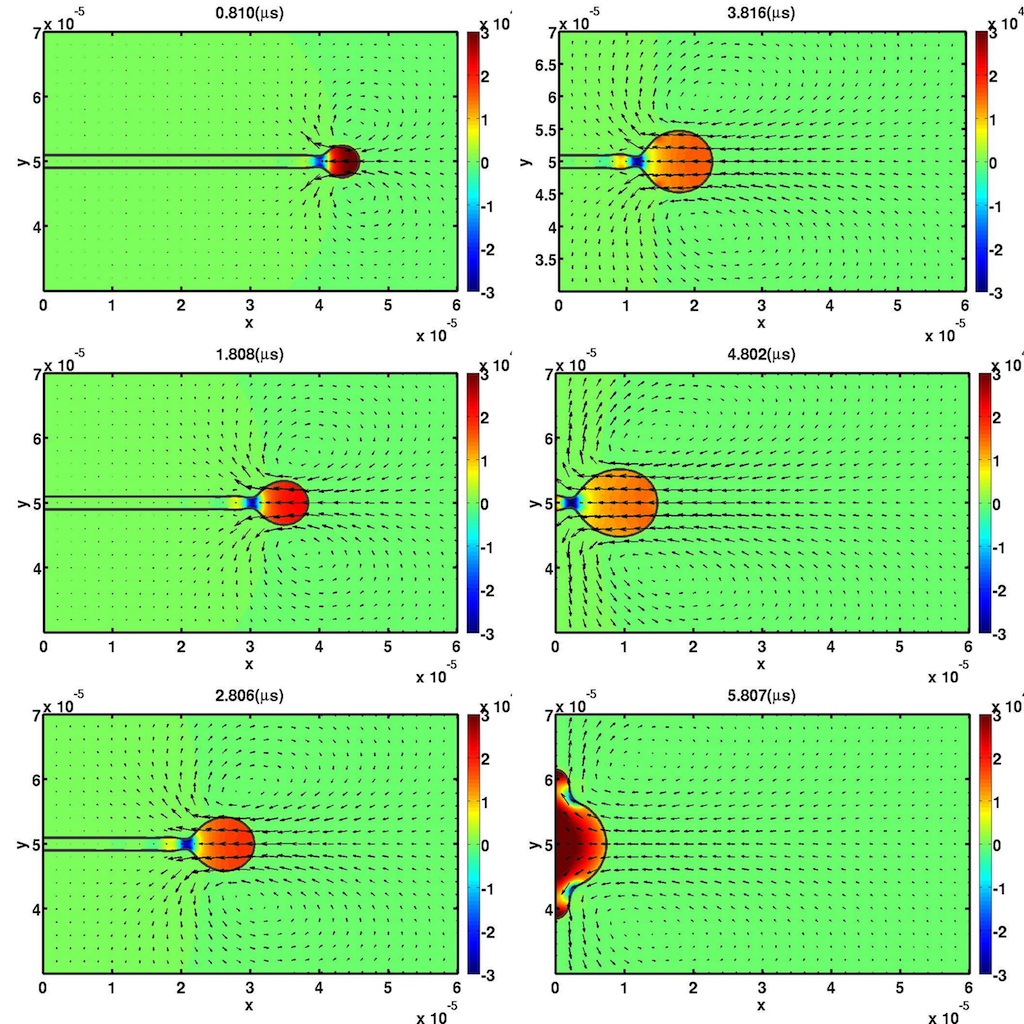}
\caption{ Thin water film retracting in air. Pressure is shown in color and velocity field is drawn on top. Surface tension force pulls the liquid film to the left.} 
\label{2D_liquid_film}
\end{figure}

In our 2D simulations, we initiate the simulations with a thin and long film closed on one side with $\vec{u}=0$ everywhere. The numerical conditions imposed on all exterior boundaries are slip boundary conditions as the domain is sufficiently far away from the long film and its retracting edge. Figure \ref{2D_gas_film} shows a thin gas film with $h=2\mu m$ retracting in a 2D domain. This 2D film does not leave behind any shedding bubbles, and the gas keeps accumulating near the edge as it retracts. If we switch the phases, the retraction of thin liquid film does not leave any drops behind either, as can be observed from Figure \ref{2D_liquid_film}. In these figures, the streamlines are plotted on top of pressure values. In both scenarios, the pressure gradients are much larger in the liquid phase and in fact the pressure in the gas phase is almost uniform. In the liquid phase of either scenario, ${Re}_{l}=\mathcal{O}(100)$ and pressure gradients predominantly balance inertial terms in the momentum equation. Moreover, a growing rim is formed in both cases with a narrow neck to its left. In the gas film problem (Figure \ref{2D_gas_film}), scaling arguments can be used to show that due to the high density and viscosity ratios of water to that of air, only very small pressure gradients can be sustained in the gas film. Thus, variation in the curvature of the film induces high and low pressure zones in the liquid phase which form a pair of counter-rotating vortices at the rim of the film. These vortices travel with the rim and displace the liquid from its neck (left of rim) to behind the rim (right of rim), causing the film to retract and the rim to grow as it collects air. For the liquid film retraction (Figure \ref{2D_liquid_film}), however, the high pressure in the rim drives the liquid towards the film, accumulating more liquid along its way and causing the rim to grow. 

\begin{figure}[H]
\centering
\includegraphics[width=\linewidth]{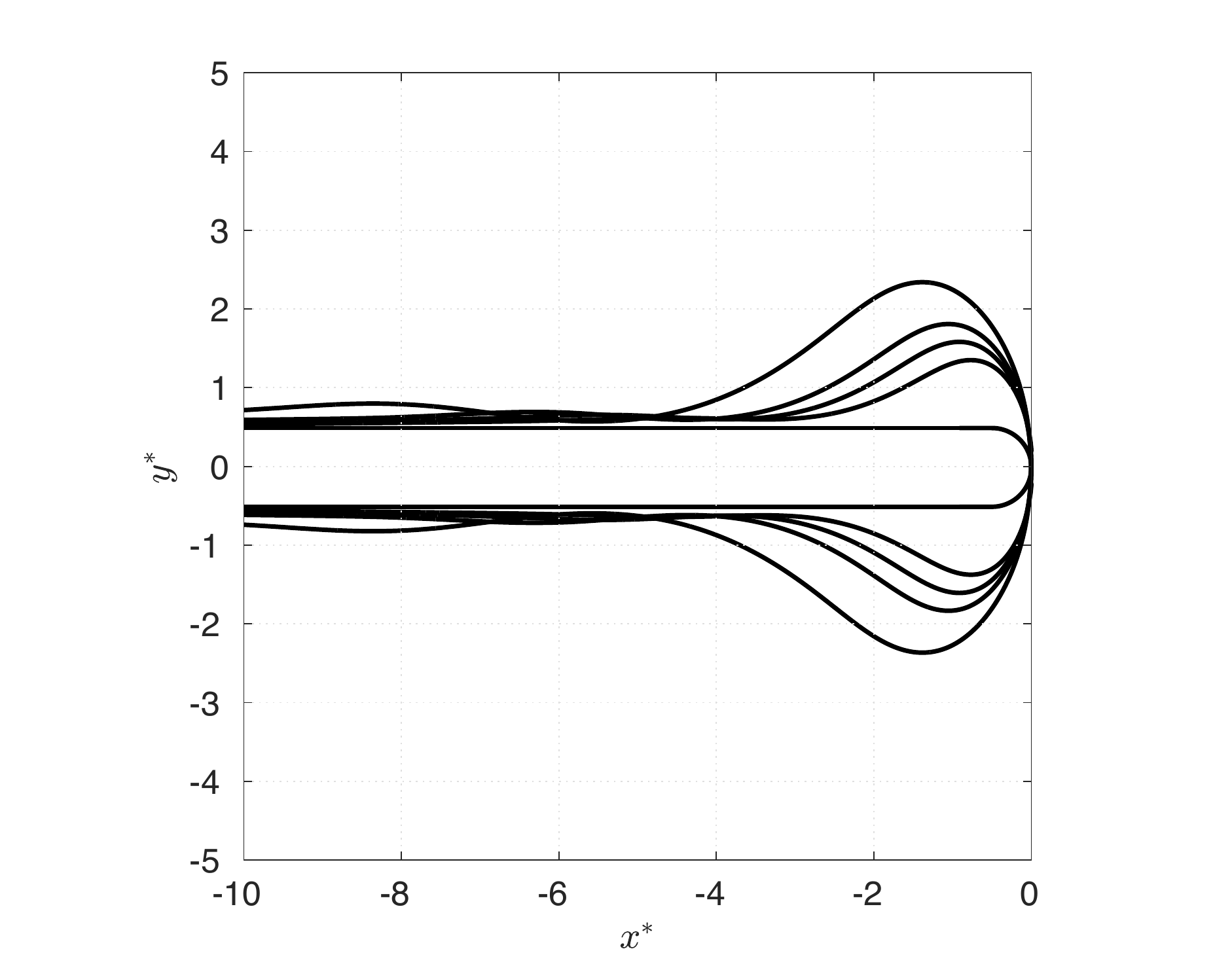}
\caption{Evolution of the thin gas film edge as it retracts plotted against dimensionless coordinates. The contours are shifted such that the rightmost point of the film is at ${x}^{*}={y}^{*}=0$.} 
\label{gas_edge_overlay}
\end{figure}

\begin{figure}[H]
\centering
\includegraphics[width=\linewidth]{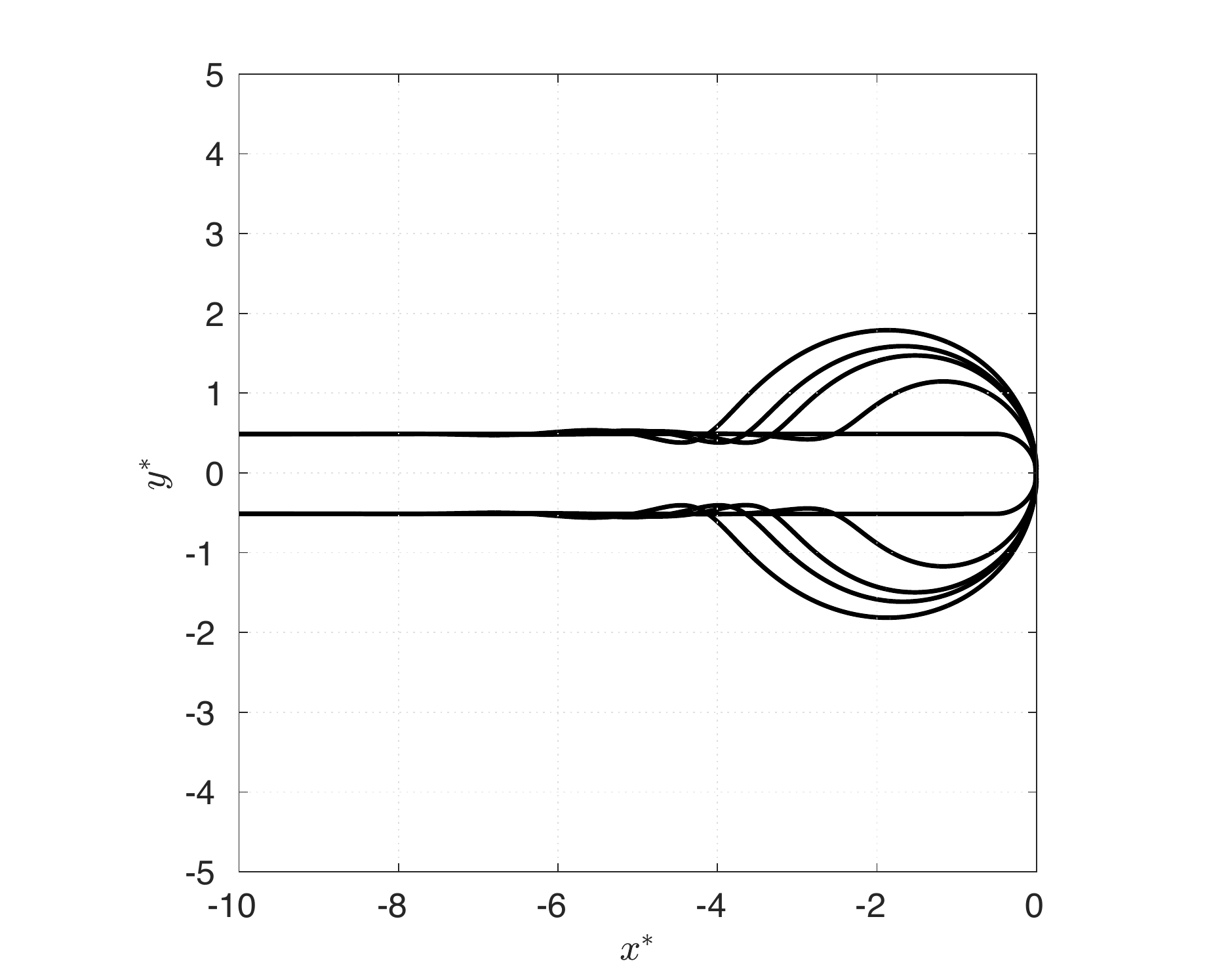}
\caption{Evolution of the thin liquid film edge as it retracts plotted against dimensionless coordinates. The contours are shifted such that the rightmost point of the film is at ${x}^{*}={y}^{*}=0$.} 
\label{liquid_edge_overlay}
\end{figure}

The edge shape evolution for gas and liquid films with $Oh=0.072$ (the films from Figures \ref{2D_gas_film} and \ref{2D_liquid_film}) are plotted against dimensionless coordinates (${x}^{*}=x/h$ and ${y}^{*}=y/h$) centered at the film tip in Figures \ref{gas_edge_overlay} and \ref{liquid_edge_overlay}, respectively.

\begin{figure}[H]
\centering
\includegraphics[width=\linewidth]{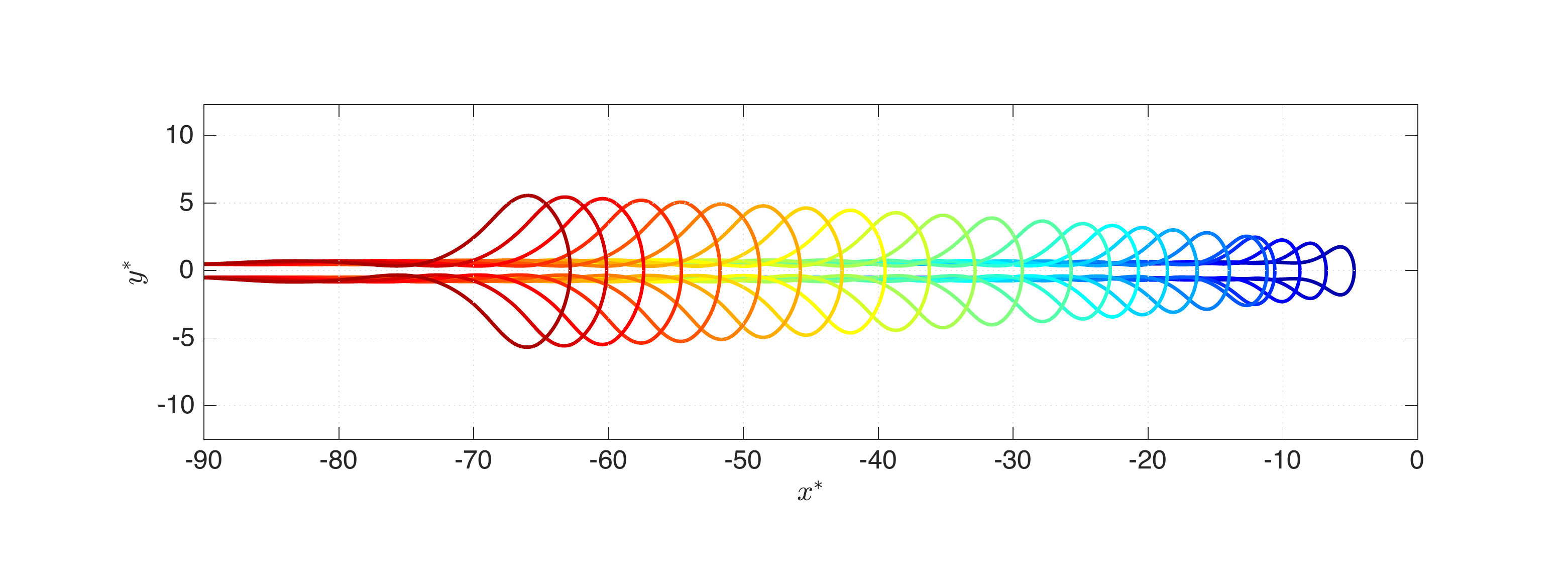}
\caption{The evolution of a thin retracting air film in background water simulated using DI in 2D. The thin film has Oh=0.072 and all lengths are normalized by film thickness.} 
\label{long_film_retraction}
\end{figure}

\begin{figure}[H]
\centering
\includegraphics[width=\linewidth]{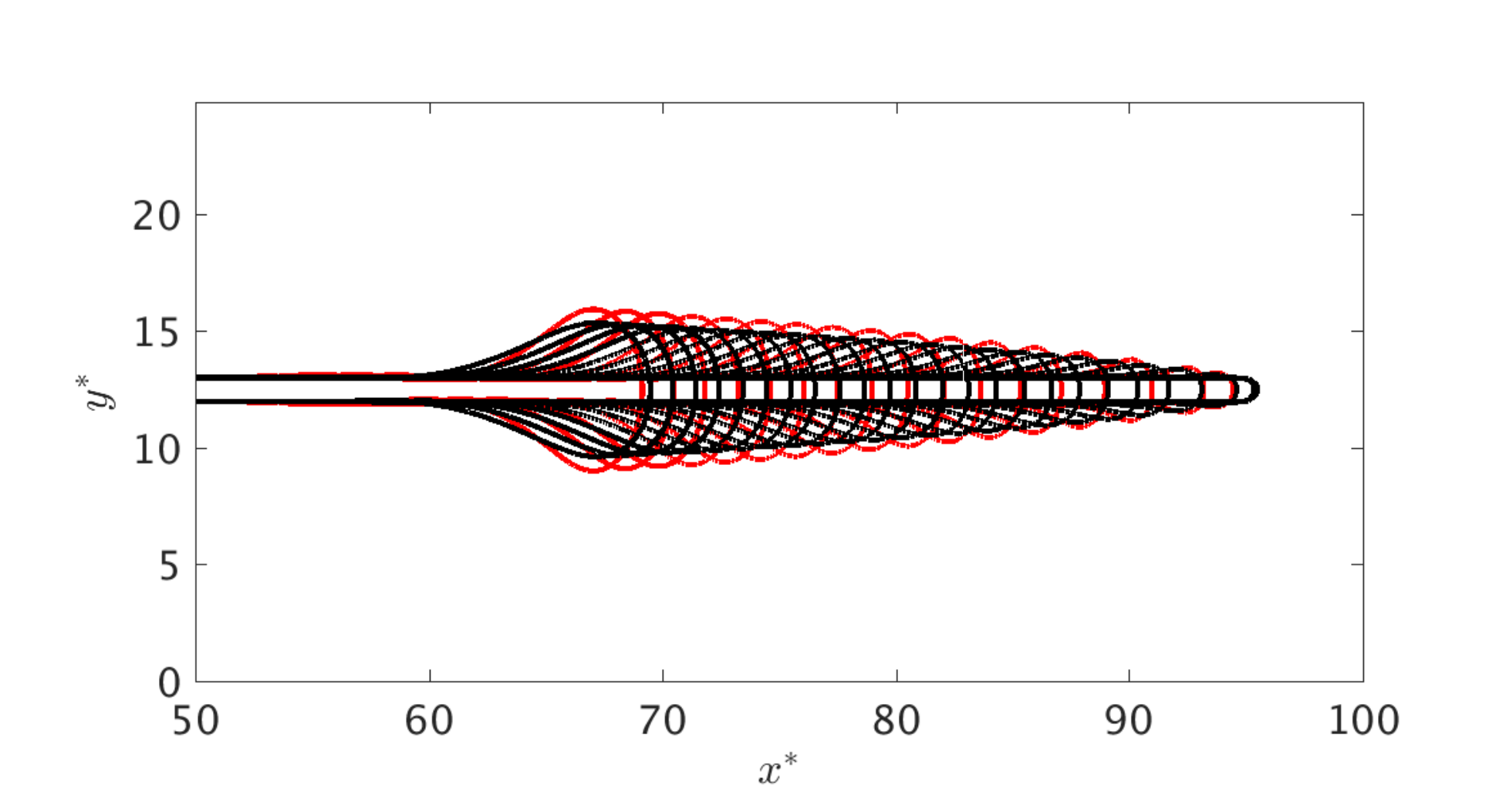}
\caption{Interfacial profile evolution of retracting thin gas films with $Oh=0.113$ and $Oh=0.320$ plotted in dimensionless coordinates in red and black respectively.} 
\label{2D_profiles_oh_113_32}
\end{figure}

 It is quite clear that the edge of the liquid film attains a much more cylindrical shape compared to the thin gas film. This observation can be intuitively explained by noting that for surface tension to create a minimal surface in the gas film case (Figure \ref{2D_gas_film}), it would need to displace much more liquid compared to the case where it only moves the flow inside the retracting liquid film (Figure \ref{2D_liquid_film}). The retraction of a very long air film in water with $Oh=0.072$ can be observed in Figure \ref{long_film_retraction}. It is clear that this film does not shed any bubbles in 2D.
Experiments (Thoroddsen \textit{et al.} 2002, Thoroddsen \textit{et al.} 2012, Tran \textit{et al.} 2013) and numerical simulations (Figure \ref{table_self_similar}) inform the relevant film thicknesses in the context of Mesler entrainment. For air/water systems, the film thicknesses are $h=\mathcal{O}(100 nm)-\mathcal{O}(1\mu m)$. As such, we consider film thicknesses of $h=100nm, 200nm, 800nm, 2\mu m$, corresponding to $Oh=0.320, 0.224, 0.113, 0.072$ respectively. Air film retraction simulations with these $Oh$ values are performed in 2D keeping resolution per film thickness the same. All simulations retracted throughout the domain without leaving any bubbles behind. The interfacial profiles of two thin films with different $Oh$, corresponding to different film thicknesses for a water-air system are shown in Figure \ref{2D_profiles_oh_113_32}. In this figure, the interface for simulations at $Oh=0.113$ and $Oh=0.320$ are plotted in red and black respectively. It seems that at lower $Oh$, where capillary effects dominate viscous effects, the rim is more cylindrical and essentially rounder. This is expected, as capillarity tends to minimize the surface of the film while retracting. Retracting liquid films (S\"{u}nderhauf \textit{et al.} 2002) exhibit a similar behavior. Further examination of this shape dependence on $Oh$ is worthwhile but out of the scope of this study.
Comparing retracting gas and liquid films, shown in Figures \ref{2D_gas_film} and \ref{2D_liquid_film} respectively, reveals the lower retraction velocity of the gas film. An analysis of momentum balance for the growing mass in the edge by the surface tension force reveals that for $Oh<1$, liquid film retraction speed is given by the Taylor-Culick velocity (Taylor 1959, Culick 1960)
\begin{equation}
{V}_{TC}=\sqrt{\frac{2\sigma}{{\rho}_{l}h}}.
\label{taylor_culick}
\end{equation}

On the other hand, the retraction speed of the edge of a gas film in a background liquid is not as simple to derive analytically. Oguz \& Prosperetti (1989) numerically showed that a relation similar to \ref{taylor_culick} holds for the retraction speed of the edge of thin gas films,
\begin{equation}
{V}_{OP}=C\sqrt{\frac{\sigma}{{\rho}_{l}h}},
\label{oguz_prosperetti}
\end{equation}
where $C$ is a constant taking a value of approximately $1$. Experiments by Thoroddsen \textit{et al.} (2002) confirmed this relation by measuring the retraction rate of thin gas sheets in drop-pool impacts. If we denote the edge location as $R$, then dimensionless velocity and time can be defined as
\begin{equation}
{u}^{*}=\frac{dR/dt}{{V}_{OP}},
\end{equation}
and
\begin{equation}
{t}^{*}=\frac{t{V}_{OP}}{h}.
\end{equation}

\begin{figure}[H]
\centering
\includegraphics[width=\linewidth]{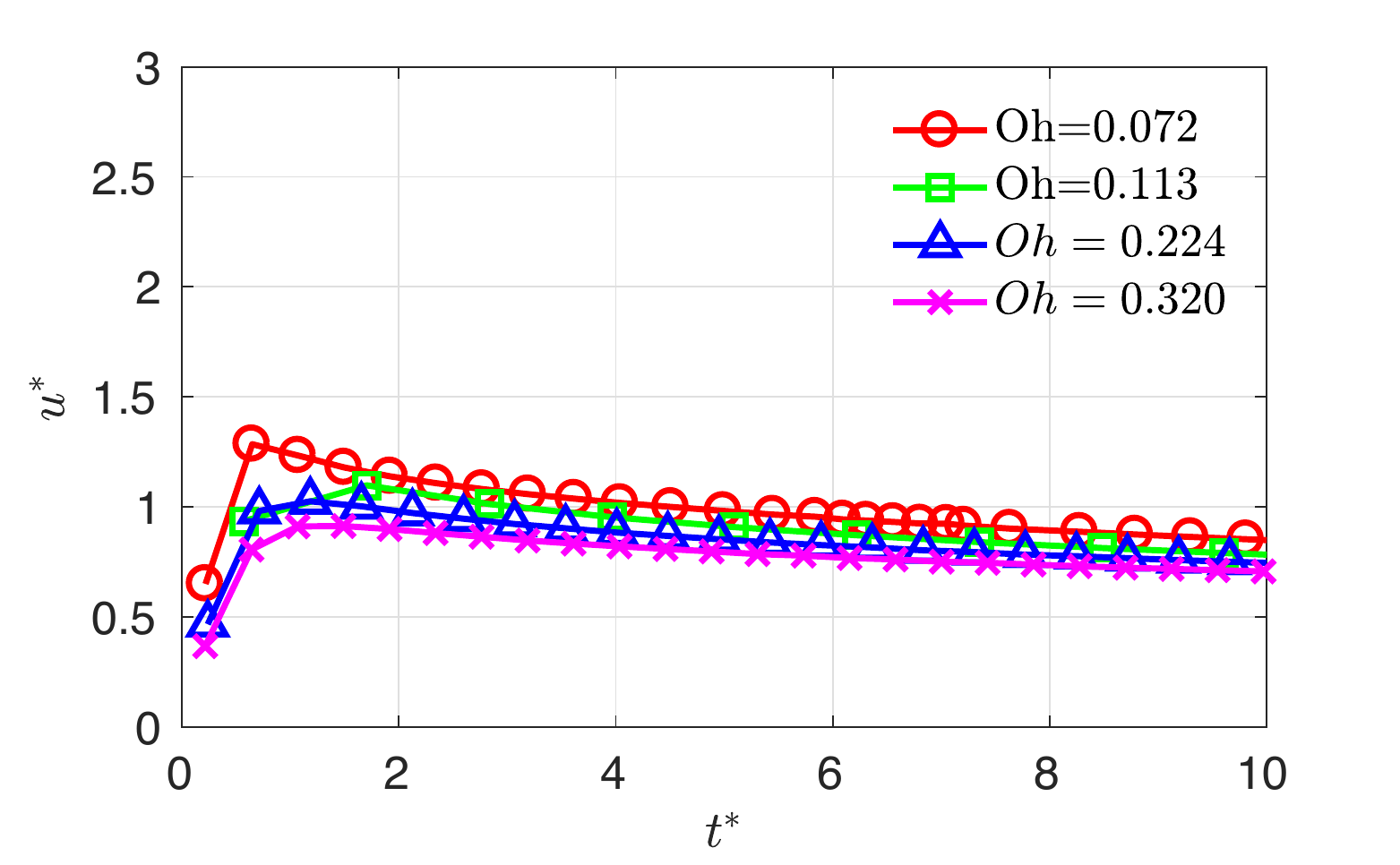}
\caption{Dimensionless retraction speed plotted against dimensionless time. ${u}^{*}\sim 1$ is equivalent to $C\sim 1$ in Equation \ref{oguz_prosperetti}.} 
\label{retraction_speed_all_Oh}
\end{figure}

\begin{figure}[H]
\centering
\includegraphics[width=\linewidth]{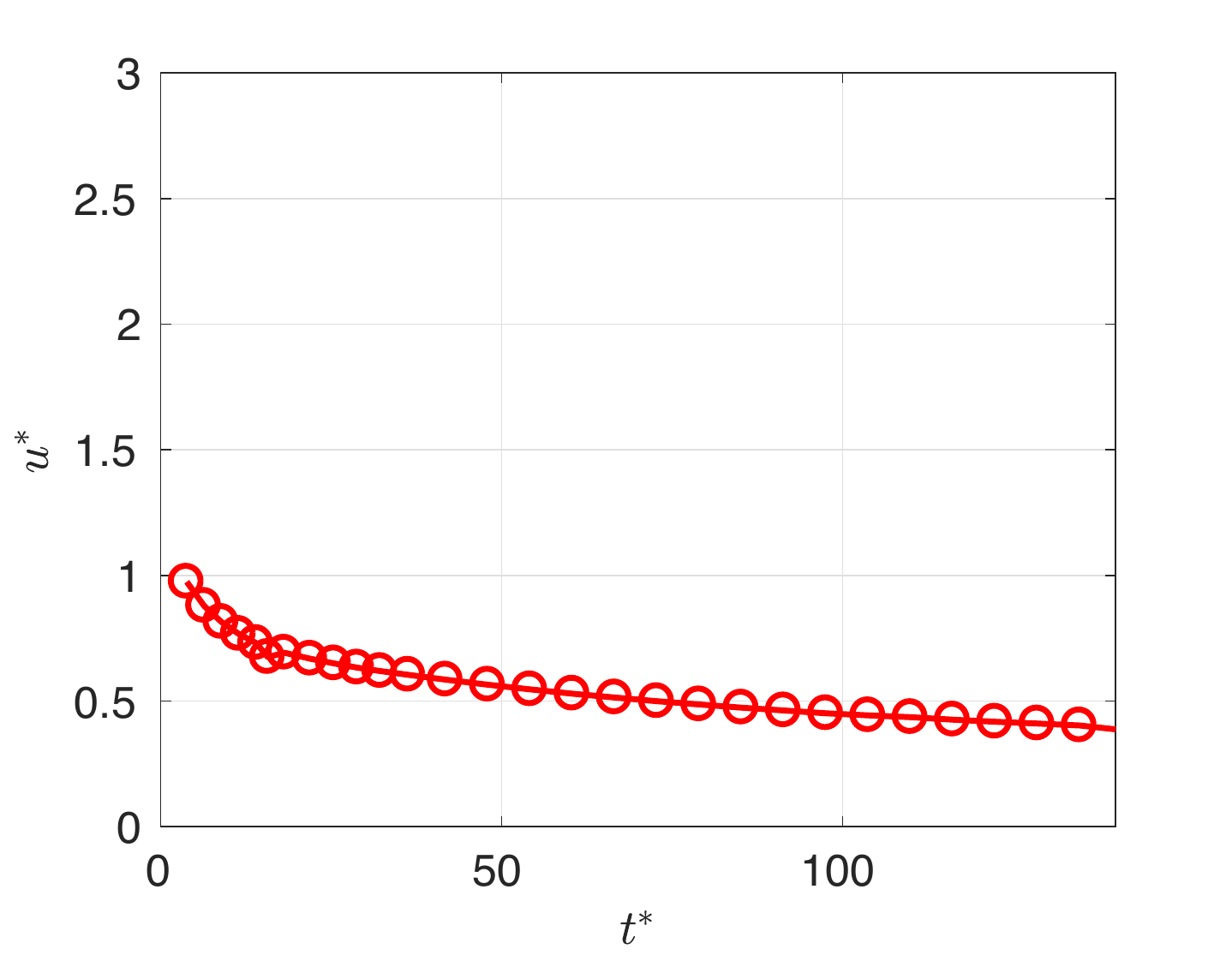}
\caption{Long-term dimensionless retraction speed plotted against dimensionless time for an Oh=0.072 air film retracting in water. ${u}^{*}\sim 1$ is equivalent to $C\sim 1$ in Equation \ref{oguz_prosperetti}.} 
\label{retraction_speed}
\end{figure}

\begin{figure}[H]
\centering
\includegraphics[width=\linewidth]{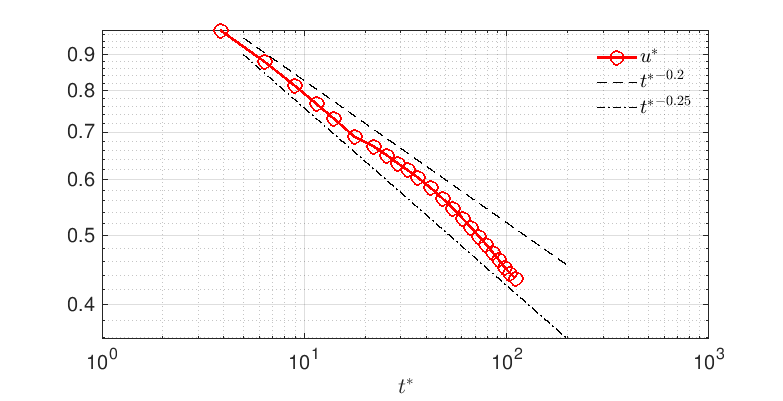}
\caption{Long-term dimensionless retraction speed plotted against dimensionless time for an Oh=0.072 air film retracting in water, demonstrating how the speed of the film edge obeys a power law in time.} 
\label{retraction_speed_power_law}
\end{figure}

Figure \ref{retraction_speed_all_Oh} shows the normalized dimensionless retraction speeds against time for different thin gas films. From this figure, the normalized retraction speeds (${u}^{*}=(dR/dt)/{V}_{OP}$) for different cases seem to be relatively similar and slowly varying. In fact, after an earlier transience the retraction speed is approximated by \ref{oguz_prosperetti} with $C\sim 1$. Thus, a thin gas film retracting in a liquid background retracts about $\sqrt{2}$ times slower than a thin liquid film with the same thickness retracting in the gas. The minor effect of $Oh$ number can be used in conjunction with experimental images to estimate the gas film thickness at rupture time. For a pair of liquid-gas working fluids, the film thickness is the only parameter that influences the film retraction dynamics. Considering Figure \ref{retraction_speed_all_Oh}, giving the scaling of ${u}^{*}\approx 1$, the film thickness can be found to be
\begin{equation}
h\approx \frac{\sigma}{{\rho}_{l}{(dR/dt)}^{2}},
\label{using_op}
\end{equation}
where the retraction rate ($dR/dt$) can be measured experimentally. This relationship is only valid for films where viscous effects are not dominant. In other words, as long as $Oh<1$, we can use Equations \ref{oguz_prosperetti} and \ref{using_op}. We will expand upon experimental techniques for measuring film thickness in the next section. A more careful inspection shows that there is a slight decrease in ${u}^{*}$ as the $Oh$ number is increased. This can be justified by acknowledging that higher $Oh$ values result in higher viscous dissipation, reducing the total energy of the system. As mentioned above, Equations \ref{taylor_culick} and \ref{oguz_prosperetti} were both derived while neglecting viscous effects. In fact, it has been shown in literature that liquid film retraction rates normalized by ${V}_{TC}$ from Equation \ref{taylor_culick} also decrease as $Oh$ is increased (S\"{u}nderhauf \textit{et al.} 2002).
While \ref{oguz_prosperetti} gives a good approximation on the retraction speed for films with $Oh<1$, the long term evolution of the retraction speed of these thin films sheds light on an important distinction between retracting thin gas films and liquid films. Figure \ref{retraction_speed} clearly shows that after an initial transitory phase, the edge velocity slowly decreases with time. This observation can be justified using conservation of momentum. For a control volume containing the edge and large enough that liquid and gas velocities vanish at the boundaries, external forces must be balanced by the rate of change of fluid momentum. For large density ratios, the inertia of the gas inside the film can be ignored. Thus, the drag force on the film which is equal (with opposite sign) to the rate of change of momentum of the liquid must balance the surface tension forces acting at the surface of the control volume. After the edge grows to become much larger than the film thickness, we can assume that the drag force on the film is dominated by the drag on its edge. The $Re$ numbers of the flow around the edge of the films of interest to us are $\mathcal{O}(100)$, and as a result the drag on the nearly cylindrical edge of these thin films can be written as ${\rho}_{l}H{(dR/dt)}^{2}{C}_{D}$ per unit length, where H is the edge diameter and ${C}_{D}$ can be approximated as a constant at such $Re$ numbers. Equating this with the surface tension force, and using mass conservation, 
$H(dH/dt)\sim (dR/dt)h$, we obtain $dR/dt \sim {t}^{-1/5}$ and ${H}\sim {t}^{2/5}$. This power law is verified in Figure \ref{retraction_speed_power_law} and a similar observation was made in experiments on anti-bubbles by \cite{Sobyanin2015}. We later show that this inverse relationship between retraction speed and edge diameter plays an important role in the 3D mechanism underlying micro-bubble shedding from gas films.  

\begin{figure}[H]
\centering
\includegraphics[width=\linewidth]{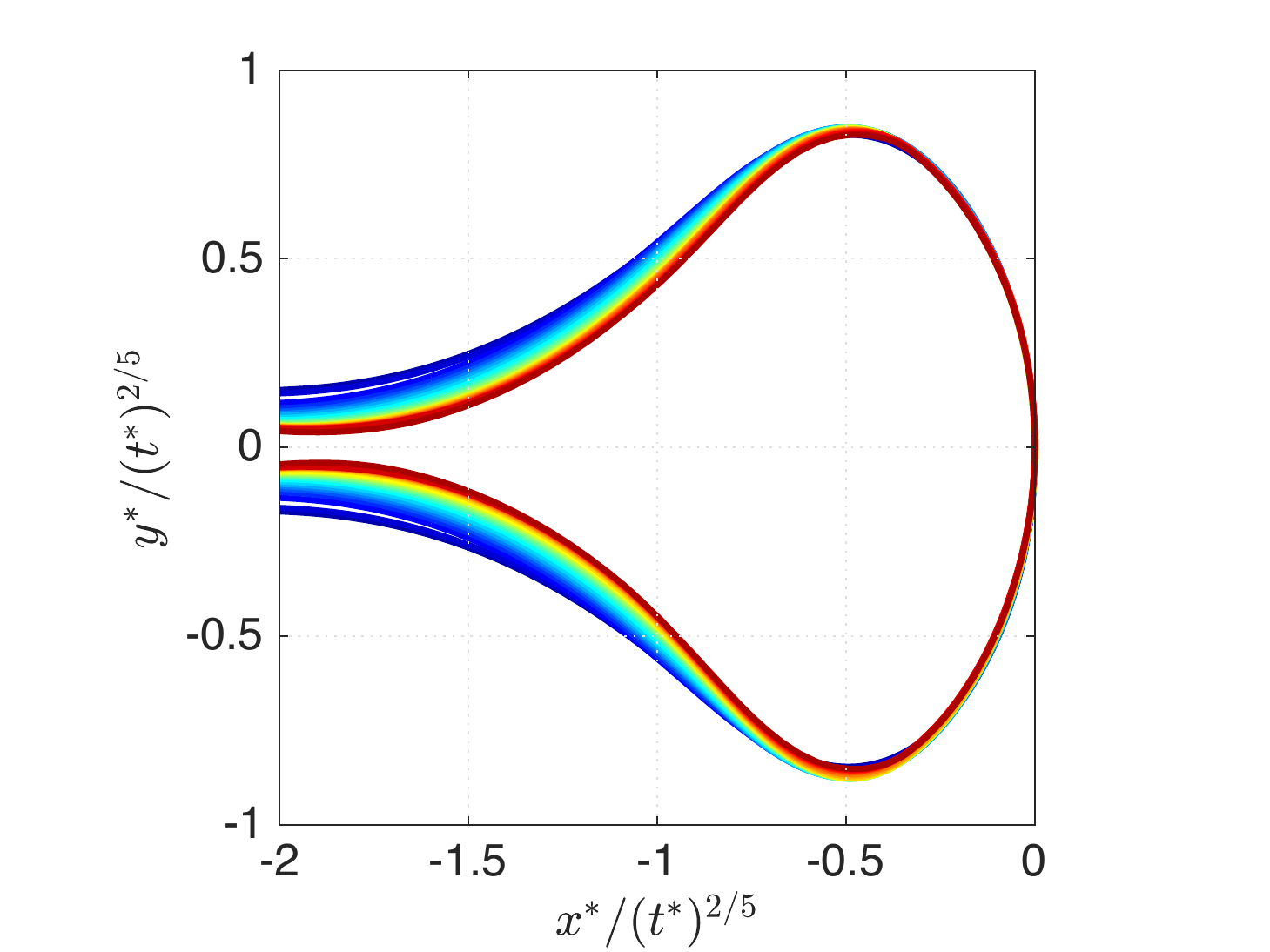}
\caption{The evolution of a thin retracting air film in background water with Oh=0.072 simulated using DI in 2D. The edge profiles shown in Figure \ref{long_film_retraction} are shifted and scaled according to the power law dictating the growth of $H$.} 
\label{self_similar}
\end{figure}

Considering the 2D retracting air films, there are no significant qualitative differences between different $Oh$ cases. The rim shape and normalized retraction speeds are similar in all $Oh$ number simulations. More importantly, none of the cases we examined in our 2D calculations resulted in micro-bubbles, while experiments on Mesler entrainment have shown micro-bubbles shedding from gas films with similar $Oh$ values. In fact, we observe that the rim finds a self-similar profile as shown in Figure \ref{self_similar}. In this figure, the edge profiles shown in Figure \ref{long_film_retraction} are shifted and then scaled based on the time in accordance with the scaling law found for the growth of the rim thickness, $H$. This confirms that 3D instabilities in the span-wise direction must be responsible for micro-bubble generation, as was also illustrated by Thoroddsen \textit{et al.} (2012). In that work, "sawtooth" instabilities were found to cause micro-bubble shedding. We will study such instabilities in the following sections. 

Based on the geometrical similarities between the results of 2D simulations of gas films at various $Oh$ numbers, it is justified to use a common mesh for 3D calculations at different $Oh$ values. From our 2D simulations, after mesh refinement, we observe that the film thickness can be resolved by 20 cells. Isotropic uniform meshes around the interfacial regions are also found to yield best results. We take advantage of these cost-effective learnings in our high-fidelity 3D simulations, presented in the proceeding sections of this paper.

\section{3D retracting thin gas films}

As explained in the previous section, our 2D simulations result in stable self-similar growing rims, indicating that 3D mechanisms are in play for break-up of micro-bubbles. In the context of drop-pool impact experiments, Thoroddsen \textit{et al.} (2002) and
Thoroddsen \textit{et al.} (2012) observed instabilities in the span-wise direction of retracting thin gas films. Such instabilities seemingly triggered micro-bubble shedding from the thinner, higher aspect ratio gas films in Thoroddsen \textit{et al.} (2012). In Thoroddsen \textit{et al.} (2002), the wave length of the undulations in the span-wise direction were found to be consistent with the maximum growth wavenumber from Rayleigh-Plateau instability. Indeed, in this section we will theoretically explain and numerically confirm that a Rayleigh-Plateau type instability in the span-wise direction of the rim of a high aspect ratio (thin and long) retracting gas film can induce micro-bubble shedding. 

Rayleigh-Plateau (RP) instability is an axisymmetric mechanism that breaks cylindrical features like liquid jets into drops with lower total free surface energy (Chandrasekhar, 2013). This capillary instability can also break up hollow cylinders into bubbles. In practice, the "hollow" cylinder can be occupied by a very low inertia fluid like a gas. As the thin gas film in our problem retracts, the size of the rim increases by accumulation of the gas volume. Overtime, the rim starts to resemble a hollow cylinder, vulnerable to RP instability. The inviscid version of this instability can be applied here as per Chandrasekhar (2013). For a hollow cylinder with radius ${R}_{cyl}$, the mode of maximum instability occurs at 
\begin{equation}
{\lambda}_{RP}=\frac{2\pi}{0.484}{R}_{cyl},
\label{wave_length_RP}
\end{equation}
and the growth rate is given by 
\begin{equation}
{\omega}_{RP}=0.820\sqrt{\sigma/({\rho}_{l}{R}_{cyl}^3)}.
\label{RP_growth_rate}
\end{equation}
These relations can provide experimentalists with a simple method, requiring just a few images, for computing the thickness of retracting thin gas films, for instance in drop-pool experiments. Measuring the thickness of the thin gas films entrapped during Mesler entrainment has been found to be rather cumbersome. Thoroddsen \textit{et al.} (2012) relied on images of the micro-bubble cloud after retraction terminated as a direct way of estimating the air film thickness at rupture. Alternatively, Tran \textit{et al.} (2013) used a technique based on colour interferometry to extract the thickness of the film. Their method, however, was limited to a narrow region around the center of the drop due to limited field of view. We have already mentioned that for $Oh<1$, a practical way to estimate film thickness is based on the retraction speed and Equation \ref{oguz_prosperetti}. This method also fails for viscous liquids, where $Oh\geq1$. Equation \ref{wave_length_RP} has a weak dependence on viscosity of the liquid, which allows the following method to be also applicable to more viscous liquids such as the ones used in the experiments of Thoroddsen \textit{et al.} (2012) and Tran \textit{et al.} (2013). The steps for our proposed method are:
\begin{itemize}
\item Measure wavelength of edge instability, ${\lambda}_{exp}$, from the experimental images 
\item Use Rayleigh-Plateau most unstable wavelength from Equation \ref{wave_length_RP} to estimate the effective diameter of the rim, ${H}_{exp}=0.484{\lambda}_{exp}/\pi$.
\item Measure distance of edge from rupture point, ${\Delta}_{exp}$ 
\item Use conservation of mass to obtain the average film thickness, ${h}_{exp}=\pi{H}_{exp}^{2}/(2{\Delta}_{exp})$.
\end{itemize}
Applying this method to the images provided by Thoroddsen \textit{et al.} (2012), results in thin film thickness of about $120nm$, which is close to the estimates provided by the authors. 

\begin{figure}[H]
\centering
\includegraphics[width=\linewidth]{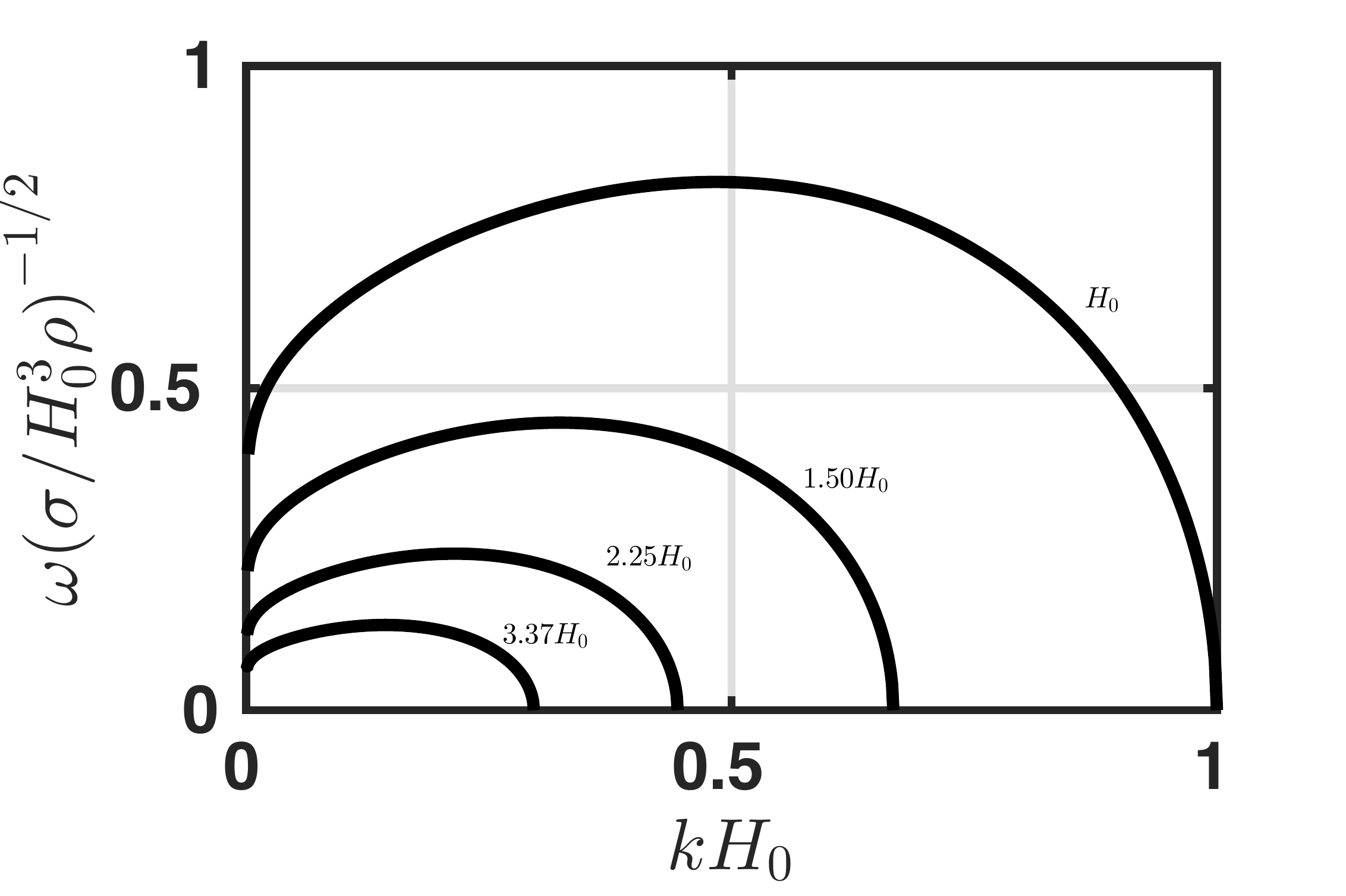}
\caption{Growth rate of RP instability at unstable wave numbers plotted for a growing rim. The curves move to the left of the plot as time progresses and the rim increases in size.} 
\label{wave_growth_time}
\end{figure}

As the film retracts, the rim continually receives air and grows. From Equations \ref{wave_length_RP} and \ref{RP_growth_rate} it is quite clear that as the rim grows, the most critical wavelength increases and the growth rate decreases. Figure \ref{wave_growth_time} shows the evolution of the unstable RP modes and their growth rate as the rim grows in time. The growing rim becomes unstable to perturbations at larger wavelengths with smaller growth rates. Therefore, growth of the rim which happens on the timescale of film retraction can stabilize Rayleigh-Plateau instability. In particular, two time scales are in competition: the time
scale associated with growth of the rim radius (${T}_{ret}$), and the RP instability in the span-wise direction (${T}_{RP}$). Bubbles can only be shed when the RP instability is much faster than the edge growth, ${T}_{RP}\ll{T}_{ret}$. If we denote the "diameter" of the rim at a given time to be given by $H$, then we have
\begin{equation}
{T}_{RP}\sim \sqrt{\frac{\rho{H}^{3}}{\sigma}}.
\end{equation}
For films with $Oh<1$, corresponding to film thicknesses of about $100nm$ and above for water-air systems, the retraction speed can be approximated to be of the order of ${V}_{OP}$ from Equation \ref{oguz_prosperetti}, by which we are ignoring the weak time-dependence of retraction velocity. Additionally, conservation of mass dictates the scaling $H dH/dt\sim h{V}_{OP}$, yielding the time scale of retraction defined as ${T}_{ret}=H/(dH/dt)$ to be
\begin{equation}
{T}_{ret}\sim \sqrt{\frac{\rho{H}^{4}}{\sigma h}}.
\end{equation}
For ${T}_{RP}\ll{T}_{ret}$ to hold, the thin film must satisfy $\sqrt{h/H}\ll1$. For a film with retraction length L, given ${H}^{2}\sim hL$, once can write this condition as:
\begin{equation}
\sqrt[4]{h/L}\ll1.
\label{requirement_inviscid}
\end{equation}
From this observation, we conclude that for films with $Oh<1$, micro-bubble shedding is only possible for films with very high aspect ratios such that even the fourth root of the aspect ratio is a small number. This explains why for disk type films such as the one shown in Figure 2.b of Thoroddsen \textit{et al.} (2002) displaying the outcome of a water drop impacting with $D=4.24mm$ and $U=0.49m/s$, azimuthal undulations stabilize and shed no bubbles. On the other hand, Figure 8.a of the same paper, shows that a thinner and longer film obtained for similar impact parameters ($D=4.24mm$ and $U=0.6m/s$), results in hundreds of micro-bubbles. This is in accordance with our scaling analysis since in the latter case a much longer and thinner film was generated before the two liquid bodies made contact.  
While we do not concern ourselves with more viscous fluids such as the silicon oils frequently used in Mesler entrainment experiments (Saylor \& Bounds 2012, Thoroddsen \textit{et al.} 2012, Tran \textit{et al.} 2013), we briefly note that in such cases the thin gas film retracts with the capillary-viscous velocity given by ${V}_{cap-visc}=\sigma/{\mu}_{l}$ (Reyssat \& Quere 2006, Aryafar \& Kavehpour 2008, Thoroddsen \textit{et al.} 2012) which is independent of the film thickness and slower than ${V}_{OP}$ (Equation \ref{oguz_prosperetti}. This increases the retraction time scale (${T}_{ret}$), while not affecting the scaling of ${T}_{RP}$, since viscous effects at such $Oh$ numbers just slightly decrease the constant behind the maximum growth rates (Ashgriz, 2011). In order to have ${T}_{RP}\ll{T}_{ret}$ it suffices to have 
\begin{equation}
\frac{1}{Oh}\sqrt[4]{\frac{h}{L}}\ll1,
\label{requirement_viscous}
\end{equation}
where of course $Oh>1$. Comparing Equation \ref{requirement_inviscid} and \ref{requirement_viscous}, one can recognize that increasing the viscosity of the liquid phase can enhance the likelihood of the RP instability leading to micro-bubble shedding. 
\subsection{3D Simulations}
We take advantage of high resolution 3D simulations of retracting thin gas films to verify the formation of edge (rim) instabilities, and study how they eventually lead to micro-bubbles. By varying $Oh$ number of the films, which is the only dimensionless parameter controlling this problem for a water/air system, we explore how the film thickness affects the dynamics and outcome of this process.

For our 3D simulations, we perform two types of simulations. First, we use a static frame of reference with a very long film, as shown in Figure \ref{sample_image}. In this simulation the edge moves towards the negative $x$ axis, until it almost reaches the boundary. In the second type of simulation which runs for a much longer time and is cheaper, a smaller length is simulated as the domain moves with the edge, allowing for the edge to remain within the bounds at all times. This is made possible by the approximately constant retraction velocity, ${V}_{OP}$, given by Equation \ref{oguz_prosperetti}. We must note that the shedding of micro-bubbles does cause the retraction speed to drift from ${V}_{OP}$, for which we account for by adjusting the frame of reference speed accordingly. We explore the role of $Oh$ using the second type of simulation.

\subsection{Static Domain Simulations}
\label{SECstaticdomain}
\begin{figure}[H]
\centering
\includegraphics[width=0.9\linewidth]{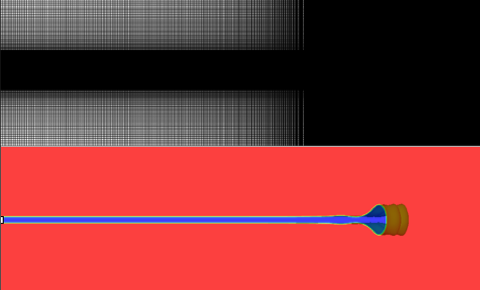}
\caption{The mesh is shown at the beginning of the static domain simulation, stretched to capture the rim dynamics accurately (top). Initial film profile obtained from extruding 2D results and shifting the fields in the $x$ direction is shown in the bottom.} 
\label{mesh_and_ic}
\end{figure}

For the 3D simulation using static domains, slip boundary conditions are used in the $x$ and $y$ boundaries, while employing periodic boundary conditions in the $z$ direction. The domain must be long enough in the $x$ direction to satisfy Equation \ref{requirement_inviscid}, and wide enough to capture the unstable wavelengths from Equation \ref{wave_length_RP}. Moreover, similar to our 2D simulations, the domain height in the y-direction is kept large enough so as to not influence the dynamics of the thin film. To be specific, for a water film retracting in air with $h=2\mu m$, which we will denote as the nominal case, we use ${L}_{x}=200\mu m$, ${L}_{y}=50\mu m$ and ${L}_{z}=100\mu m$. Our Cartesian mesh is stretched in $x$ and $y$ such that it has the highest concentration around the edge, and resolves the film thickness with at least $20$ mesh points. The mesh at the beginning of the simulation is shown in Figure \ref{mesh_and_ic}. It must be noted that one of the downsides of this type of simulation is that the mesh has to be altered after the film has retracted past the refinement zone.
Initial phase and velocity fields for this simulation are extrusions of 2D simulation results perturbed in the spanwise direction as also shown in Figure \ref{mesh_and_ic}. These perturbations are periodic waves with random small amplitudes and random phases that shift a 2D snapshot in the $x$ direction. Once the instability is triggered, self sustaining shedding phenomena is observed over the long-term simulation. 

\begin{figure*}
\centering
\includegraphics[width=0.7\textwidth]{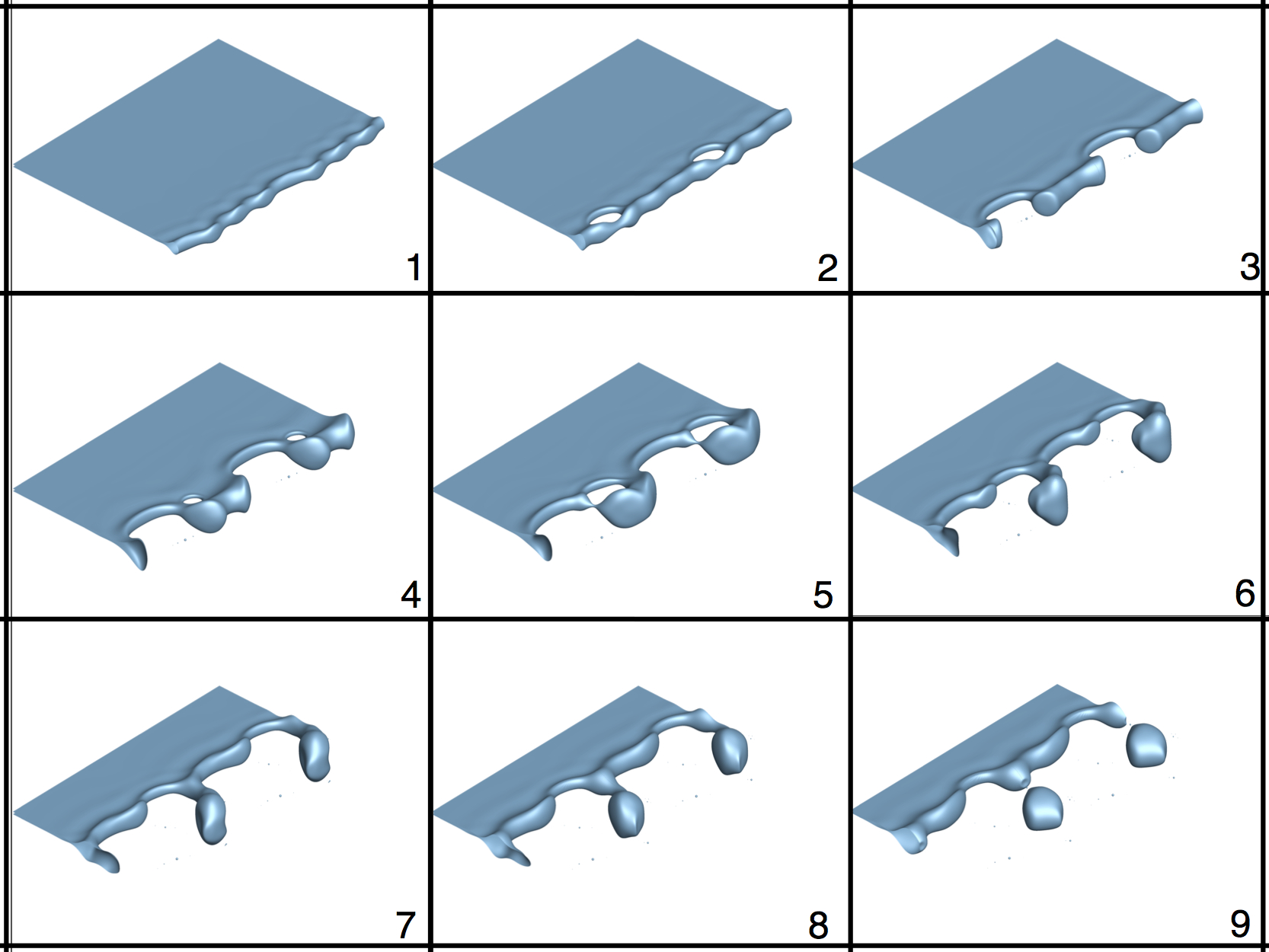}
\caption{The time-progression of static domain results for a retracting air film with $Oh=0.072$ in an initially static background of water, with the numbers indicating the order in time. Note that for visualization, the results are copied once in the z direction.} 
\label{3D_sim_type_1}
\end{figure*}

\begin{figure*}
\centering
\includegraphics[width=0.8\textwidth]{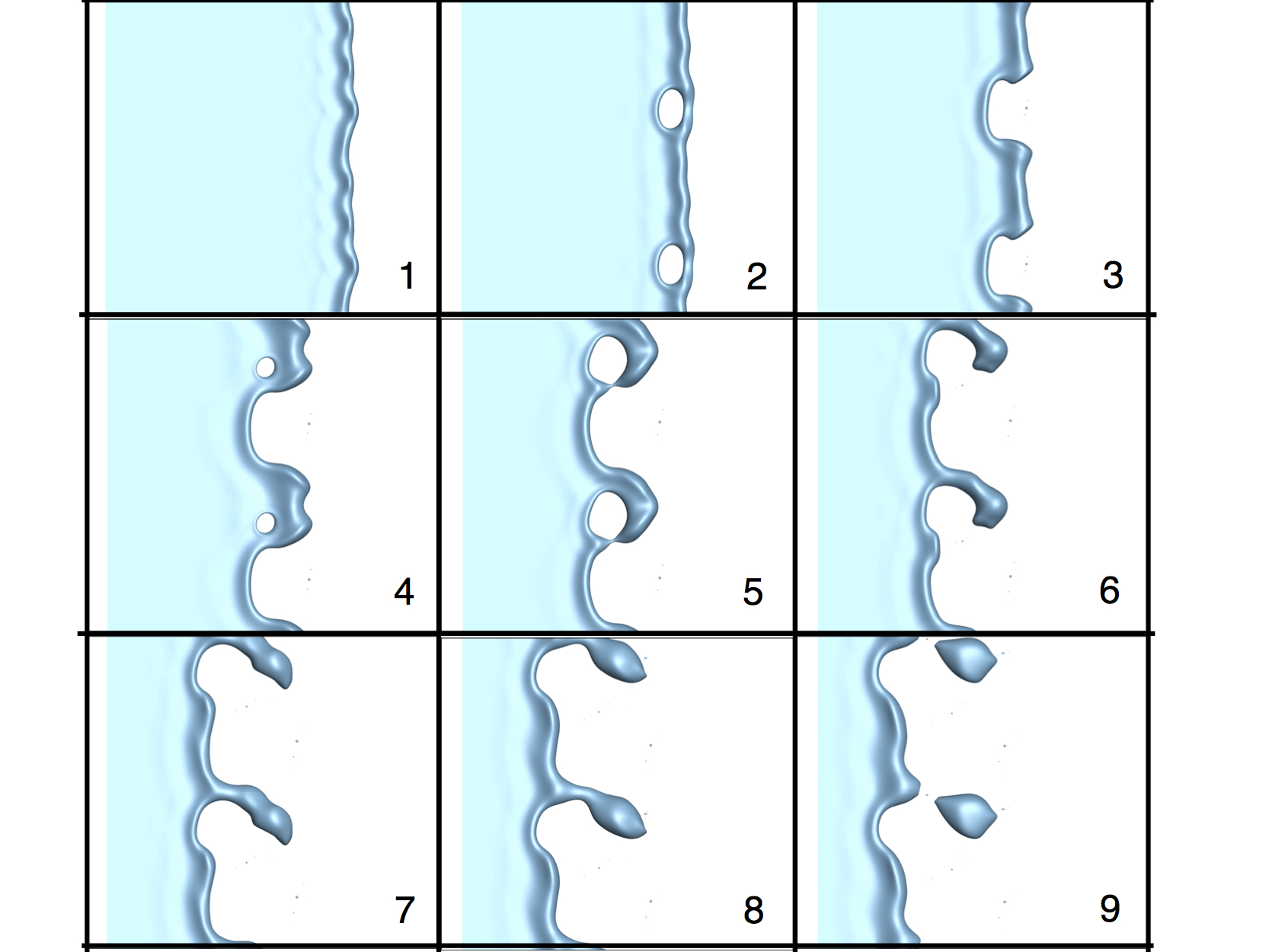}
\caption{A top view of the time-progression of static domain results for a retracting air film with $Oh=0.072$ in an initially static background of water, with the numbers indicating the order in time. Note that for visualization, the results are copied once in the z direction.}
\label{3D_sim_type_2}
\end{figure*}

The evolution of this simulation is shown in Figure \ref{3D_sim_type_1}. Figure \ref{3D_sim_type_2} also shows the $x-z$ plane view of the same images. For visualization, the simulation results are repeated in the span-wise direction in both figures. The temporal order of the panels are indicated with numbers. As shown in these figures, there are multiple steps involved in the shedding of a micro-bubble. First (panel 1,2), a span-wise edge instability causes punctures in the film that was retracting stably in 2D simulations. Next (panel 2), the hole grows such that a gas filament remains at the edge of the film. The gas filament breaks in a manner similar to primary atomization (Ashgriz, 2011), giving birth to mother and satellite bubbles (hardly visible on panel 3). The geometry of the thin film edge is severely disturbed at this stage. We will later demonstrate that this is in fact the norm for a retracting thin gas film, because once the uniform profile is broken, the retraction speed is variable for different points on the edge, depending on the local thickness. The thinner parts of the film retract with a larger speed than the thicker parts, as was explained before, leaving the larger parts of the edge behind. As gas fills the thick part of the edge another rupture occurs (panel 4), leading to another filament and its subsequent break-up (panel 5, 6). Eventually (panels 7-8), the thickest part of the edge is connected to the film via a filament that thins out as the film retracts with higher velocity until it breaks up and leaves a large micro-bubble behind (panel 9). It is clear that this mechanism is very different from how droplets pinch-off from the edge of unstable liquid sheets (Agbaglah \textit{et al.}, 2013).
\subsection{Moving Domain Simulations}
\label{SECmovingdomain}
\begin{figure*}
\centering
\includegraphics[width=0.7\textwidth]{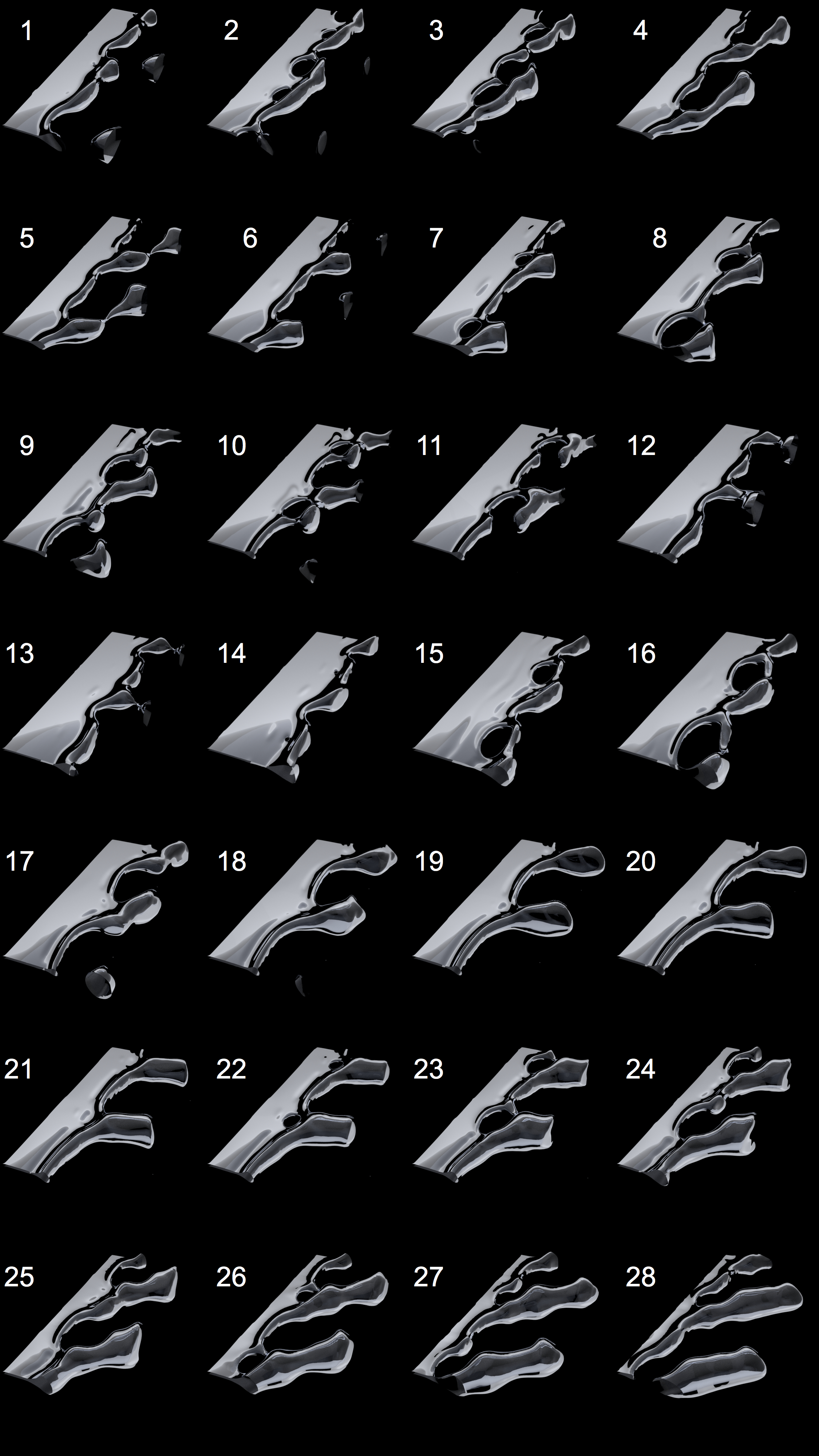}
\caption{The time-progression of moving domain results for a retracting air film with $Oh=0.072$ with the numbers indicating the order in time. Note that for visualization, the results are copied once in the z direction.} 
\label{3D_sim_mm}
\end{figure*}

The requirement for very high aspect ratio films (Equation \ref{requirement_inviscid}) renders static domain simulations expensive. Moreover, as the edge retracts, the mesh has to be modified periodically as to situate the refinement region around the edge. Given the almost constant retraction velocity of these films, it is possible to perform simulations where the frame of reference is moving with ${V}_{OP}$, given by Equation \ref{oguz_prosperetti} plus appropriate small correction as observed in Figure \ref{retraction_speed_all_Oh}. For these simulations we have inflow and outflow boundary conditions in the $x$ direction, while slip and periodic boundary conditions are imposed in $y$ and $z$, respectively. Compared to the static domain simulations, the domain has been abridged in the $x$ direction, but is long enough such that the inflow can be assumed to be given by ${V}_{OP}$ in both phases. For outflow, we use convective boundary conditions. The initial conditions are from the static domain simulations, where we continue simulating the flow around the edge after the events shown in Figure \ref{3D_sim_type_1} and \ref{3D_sim_type_2}. We performed these simulations on different $Oh$ value films, given by $Oh=0.320, 0.224, 0.113, 0.072$, while fixing the density and viscosity ratios. All simulations presented up to this point were from the nominal case, $Oh=0.072$, corresponding to a $h=2\mu m$ film. Snapshots for moving domain simulations of this nominal case are shown in Figure \ref{3D_sim_mm}. Once again, the simulation results are repeated in the $z$ direction for visualization purposes. It is clear that the mechanism for formation of micro-bubbles is further repetition of the static domain simulations. Namely, the thicker parts of the rim retract with a slower velocity than the rest of the film while continuing to grow due to accumulation of air. This relative growth of the thicker parts of the perturbed edge eventually causes puncture in the film. The hole then grows to separate one side of the thick feature. At this stage, the thick feature is connected to the film via a filament and surface tension forces act towards making the feature more spherical or bubble-like. This, in addition to the slower retraction rate of the feature thins the connecting filament until the micro-bubble is separated. In these simulations, since the domain size is smaller in the $x$ direction, we often do not capture a whole micro-bubble at its break-up time. In the following, we analyze the results of these simulations and the steps leading to micro-bubbles in detail. Finally, we conclude this section by studying the effect of varying $Oh$ number.
\subsection{Edge Instability}
\label{SECedgeinstability}
\begin{figure}[H]
\centering
\includegraphics[width=\linewidth]{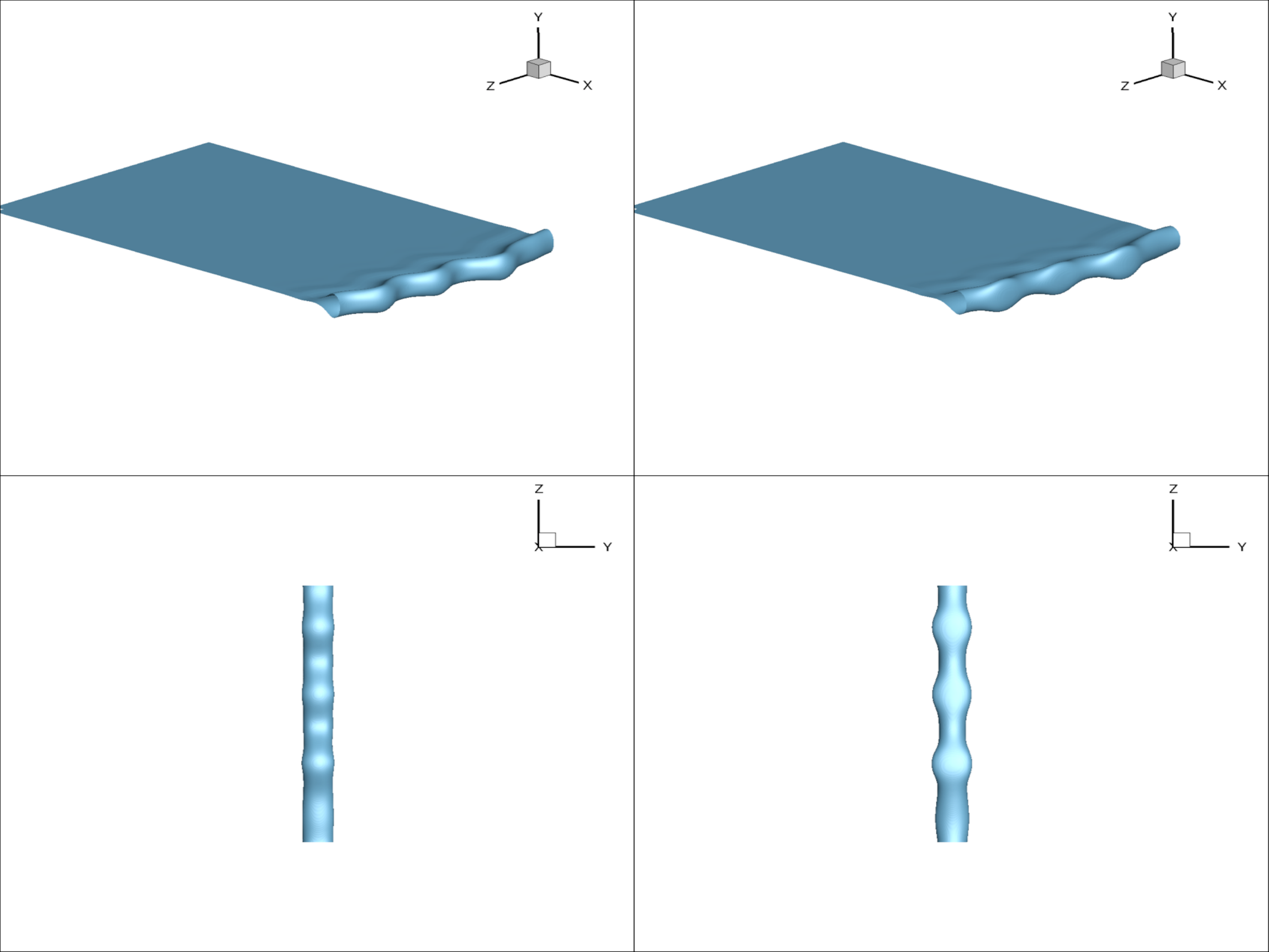}
\caption{Growth of undulations from left to right showing how the rim is susceptible to a capillary instability reminiscent of Rayleigh-Plateau instability.} 
\label{undulations}
\end{figure}

Initial conditions for the 3D simulations were obtained by extruding 2D results in the $z$ direction and perturbing them by displacing all fields in the x direction (Figure \ref{mesh_and_ic}). As the film retracts, undulations appear on the rim. Figure \ref{undulations} shows how these undulations grow for a perturbed thin air film retracting in water with $h=2\mu m$. The left and right images in Figure \ref{undulations} show before and after the growth of the undulations, respectively. The wavelength of these undulations are about $\lambda=28\mu m$. These undulations start to appear as the thickness of the rim reaches $H\approx9 \mu m$. From Rayleigh-Plateau instability, the smallest unstable wavelength is given by $\pi H\approx28\mu m$ and the maximum growth, as given by Equation \ref{wave_length_RP} happens at a wavelength of ${\pi H}/{0.484}\approx58.3\mu m$. The instability governing a growing rim connected to a thin film is different from a RP instability for a hollow cylinder. Moreover, it is difficult to establish the rim thickness when the linear phase of the instability is active. In Thoroddsen \textit{et al.} (2003), they modeled the rim as a cylinder where $H\approx 2h$ and found the measured wavelengths to match the maximum growth wavelengths for that particular rim thickness. In a similar manner, if we assume that the instability kicks in when $H\approx 2h$, the maximum growth rate would happen at a wavelength of around $26\mu m$, which is very close to the wavelength of the undulations we see in our numerical results. Regardless of these details, it is reasonable to conclude that a Rayleigh-Plateau type capillary instability is responsible for the undulations we see in Figure \ref{undulations}.
\subsection{Hole Formation}
From Figures \ref{3D_sim_type_1}, \ref{3D_sim_type_2} and \ref{3D_sim_mm}, we see that the formation of holes in the neck region of the rim is one of the necessary steps for micro-bubble shedding from thin retracting gas films. This is in contrast to how drops pinch off from unstable liquid sheets (Agbaglah \textit{et al.}, 2013) or how ligament or finger break-up leads to drops in atomization processes. Recently, Zaleski \textit{et al.} (2017) performed high fidelity numerical simulations of turbulent two-phase mixing layers and observed that the formation of holes and their rapid growth can also be responsible for generating droplets (primary atomization). For thin gas films at scales relevant to Mesler entrainment, the formation of these holes is critical in the process of micro-bubble shedding, and the thin films will remain semi-stable if such events do not happen.

\begin{figure}[H]
\centering
\includegraphics[width=\linewidth]{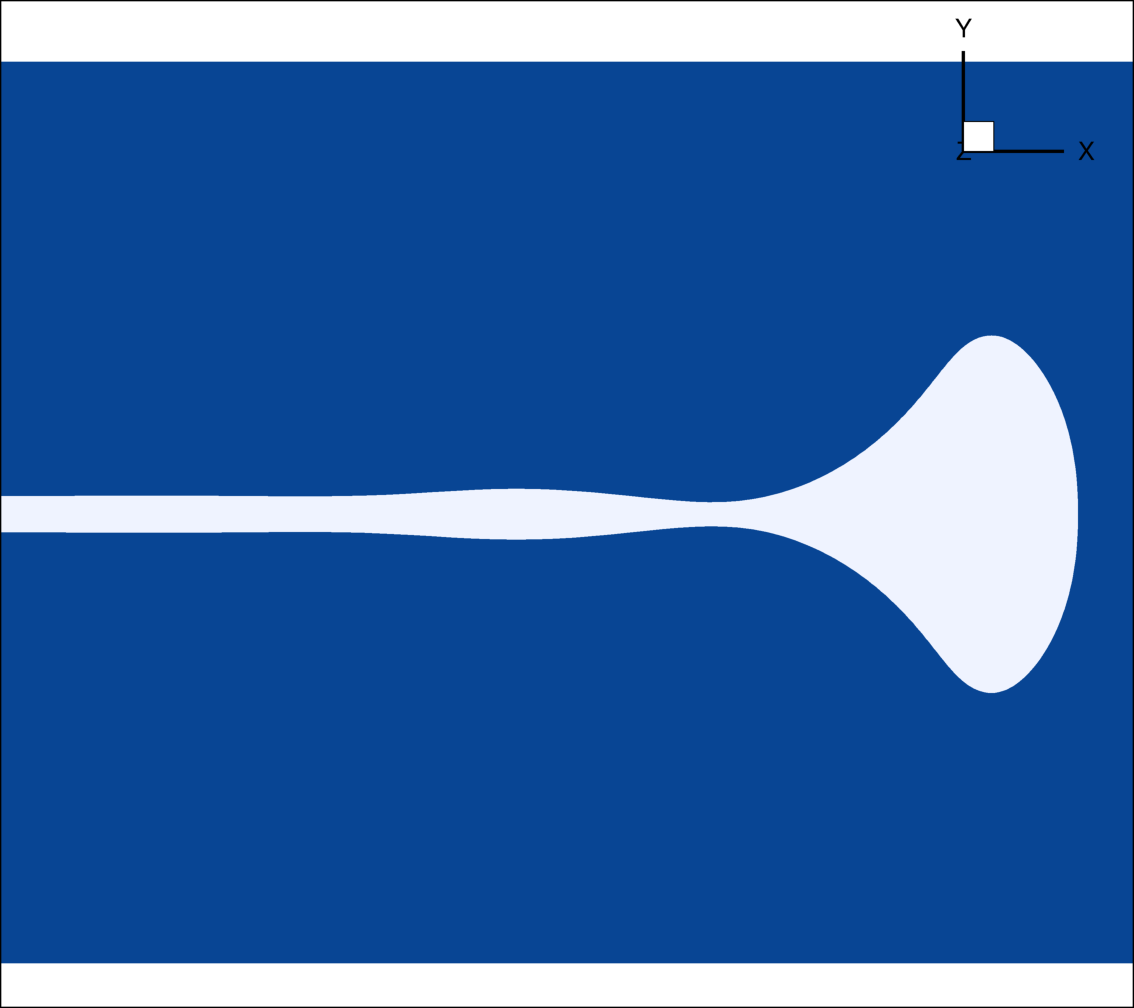}
\caption{A sample zoom-in image on the thin gas film, showing the capillary waves and thin neck region of the rim.} 
\label{capillary_wave}
\end{figure}

For small $Oh$ number films such as the nominal case studied here, capillary waves connect the thick rim to the uniform thickness film region. A prominent feature of these waves is a narrow neck region just before the rim connection to the film. This neck region, which is also observable in 2D simulations, can be seen in Figure \ref{capillary_wave} which is showing a typical $x-y$ plane of the thin film from the 3D simulations.
As the rim becomes unstable in the span-wise direction, the neck of the film at different $z$ values attains different axial positions and thicknesses. The superposition of variation of the film thickness due to capillary waves induced by the RP instability with the 2D mechanism that induces the neck results in regions with minimum film thickness that are most hospitable to puncture.
\begin{figure}[H]
\centering
\includegraphics[width=\linewidth]{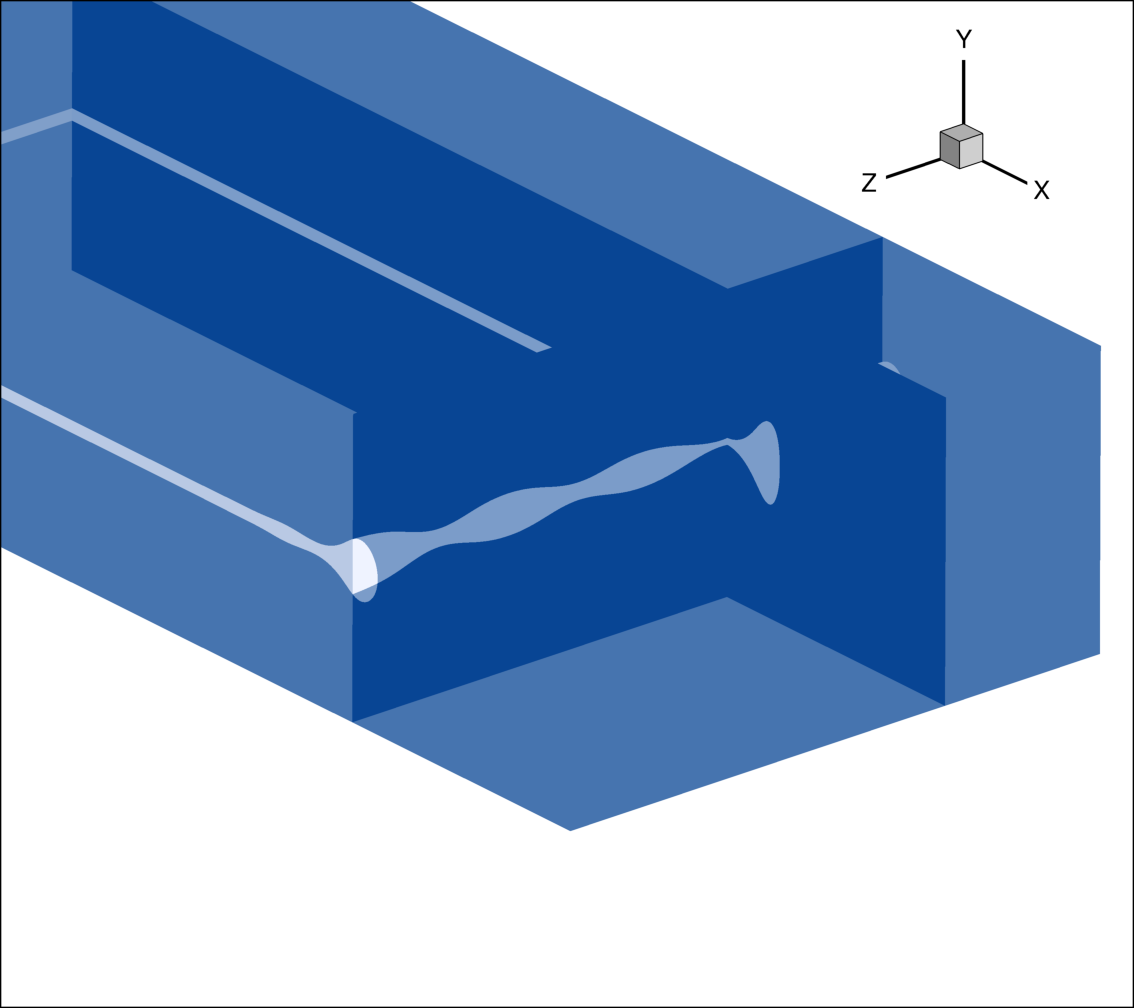}
\caption{Slices on a snapshot of the thin film prior to hole formation, showing how the interplay between the capillary waves and the film topology can lead to puncture of the film.} 
\label{capillary_wave_3D}
\end{figure}
 Figure \ref{capillary_wave_3D} shows a snapshot before a hole is formed indicating variation of film thickness on two perpendicular x-y and y-z planes. This point is where the minimum thickness of the unstable span-wise wave coincides with the minimum thickness region of the film in the stream-wise direction (the film neck), causing film rupture. After such holes are formed, they grow radially outwards until they form a filament at the edge of the film that eventually breaks, resulting in major disturbances to the shape of the rim (Figures \ref{3D_sim_type_1}, \ref{3D_sim_type_2}, \ref{3D_sim_mm}). Such disturbances cause non-uniformity in the thickness of the rim and consequent non-uniform retraction velocities that lead to micro-bubble shedding in the long run. 

\subsection{Filament Break-up}
\label{SECfilamentbreakup}
The break-up of liquid filaments plays an important role in combustion and atomization. There have been many studies focused on the non-linear mechanism leading to the formation of mother and satellite drops (Tjahjadi \textit{et al.} 1992, Ashgriz \& Mashayek 1995, Ashgriz 2011). Satellite bubbles have been observed experimentally, especially in the context of the singular process of bubble pinch off (Burton \textit{et al.} 2005, Keim \textit{et al.} 2006, Thoroddsen \textit{et al.} 2007). The dynamics of a thinning gas filament have been found to be significantly different from that of a thinning liquid filament (Burton \textit{et al.}, 2005). Thoroddsen \textit{et al.} (2007) observed the formation of a thin neck as a gas filament becomes very thin. This thin neck attained higher aspect ratios as the viscosity of the liquid was increased. In Gordillo \& Fontelos (2007), the authors performed numerical simulations of the process of separation of two identical bubbles placed in shear flow. Similar to the experiments of Thoroddsen \textit{et al.} (2007), a thin filament formed, which later turned into an elongated satellite bubble. Numerical results in Gordillo \& Fontelos (2007) also revealed that the size of the bubble decreased as the density ratio (${\rho}_{l}/{\rho}_{g}$) was increased, explaining why it is much more difficult to observe satellite bubbles compared to satellite drops. Moreover, the numerical simulations in that work showed that the ratio of the satellite bubbles to the original bubbles (mother bubbles in our case) is of the order of ${10}^{-3}$. 

\begin{figure}[H]
\centering
\includegraphics[width=\linewidth]{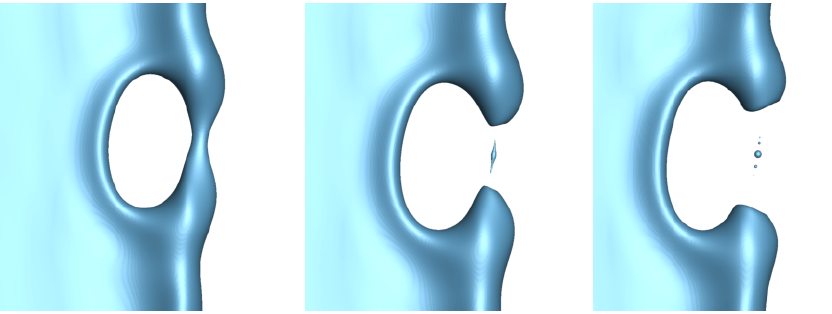}
\caption{Gas filament formation and break-up shown.} 
\label{zoom-in-filament}
\end{figure}

In our simulations, filaments are formed when growing holes in the film approach the rim (e.g. panels 2,5 and 8 in Figures \ref{3D_sim_type_1} and \ref{3D_sim_type_2}). In Figure \ref{zoom-in-filament} a zoom-in on a filament at different times is presented.
In all cases we first see an elongated bubble break from both ends, disconnecting it from the rest of the film (middle panel in Figure \ref{zoom-in-filament}). This bubble resembles the elongated drops in liquid jet break-up events (Tjahjadi \textit{et al.} 1992, Ashgriz 2011), and the elongated satellite bubble in simulations of Gordillo \& Fontelos (2007). Surface tension then, gives this bubble a spherical shape, and occasionally satellite bubbles on its sides. We find the diameter of the mother bubble (main bubble remaining from the filament) to be about $1\mu m$ for our nominal simulations ($h=2\mu m$ air film in water). These bubbles are resolved with about 10 cells but their size is not consistent with the predictions of Gordillo \& Fontelos (2007), possibly due to the low resolution in the preceding elongation stage, and also the necessity for including Van der Waals forces to capture such dynamics correctly. In any case, these bubbles are very small and their high dissolution rates render them unimportant to the Mesler entrainment process which is primarily focused on micro-bubbles in the range of $10\mu m-100\mu m$. An important feature we see in our simulation results is the formation of flattened, ragged ends on the locations where the filament was connected to the film (middle and right panels in Figure \ref{zoom-in-filament}). These features were also observed in experiments for low viscosity ambient liquids such as water (Burton \textit{et al.}, 2005, Thoroddsen \textit{et al.} 2007). This topological feature in fact plays an important role in the subsequent formation of holes and bubbles in the film as can be seen in Figure \ref{3D_sim_mm}. 

\subsection{Micro-bubble Statistics}
\label{SECmicrobubble}
The main motivation of this work is to understand how micro-bubbles are shed from thin retracting gas films and to gain a predictive capability with regards to their size and rate of production. In this vein, we examine the micro-bubbles that are separated from the thin film as shown in Figures \ref{3D_sim_type_1}, \ref{3D_sim_type_2} and \ref{3D_sim_mm}. Note that by micro-bubbles we do not refer to the very small bubbles remnant from the broken filaments, but rather the larger bubbles that have diameters of $\mathcal{O}(10\mu m)$, such as the bubble formed in panel 9 of Figures \ref{3D_sim_type_1} and \ref{3D_sim_type_2}. So far, in our simulations of the nominal case, we have seen 3 micro-bubbles shed from the thin gas film corresponding to a $h=2\mu m$ thick air film retracting in water. Snapshots showing the separation of the first bubble are provided in panel 9 of Figures \ref{3D_sim_type_1} and \ref{3D_sim_type_2}. The other two micro-bubbles are separated in panels 5 and 13 of Figure \ref{3D_sim_mm}. Table \ref{table_microbubbles} summarizes the statistics regarding these bubbles. The diameter of the bubbles (${D}_{bubb}$), in addition to the time it took them to be shed (${\Delta t}_{bubb}$) is presented in dimensional and dimensionless form in the table. Bubble diameters are obtained from the volume of the bubbles, which can be computed via calculating the volume left in the film at a given time, and by using the known rate of gas entering the domain to determine the amount of gas leaving the film in the form of micro-bubbles. 
\begin{table}[H]
\centering
\begin{tabular}{c|c|c|c|c}
Bubble  & ${D}_{bubb}(\mu m)$ & ${\Delta t}_{bubb}(\mu s)$ & ${D}_{bubb}/h$ & ${\Delta t}_{bubb}{V}_{OP}/h$ \\ \hline
1 &  26.7   &    19.9  &    13.35 &  61.4  \\
2 &  31.8   &    10.5  &    15.9  &  32.36 \\
3 &  34.2   &    20.6  &    17.1  &  63.5  \\ 
\end{tabular}
\vspace{4mm}
\caption{Bubble statistics and time intervals taken for shedding enlisted in dimensional and non-dimensional form.}
\label{table_microbubbles}
\end{table}

Based on the results so far it seems that for the nominal case ($Oh=0.072$), the diameter of the bubbles is approximately given by ${D}_{bubb}\approx15h$, as evident from the fourth column of Table \ref{table_microbubbles}. Interestingly, the interval for their shedding is not necessarily similar as can be seen from the third and fifth columns of the table. These simulations are still in progress, and by capturing more bubbles we can obtain reliable statistics, from which we can draw conclusions regarding the bubble size distribution and shedding rates from this film. 
\subsection{Dependence on Ohnesorge (Oh) Number}
\begin{figure}[H]
\centering
\includegraphics[width=0.8\linewidth]{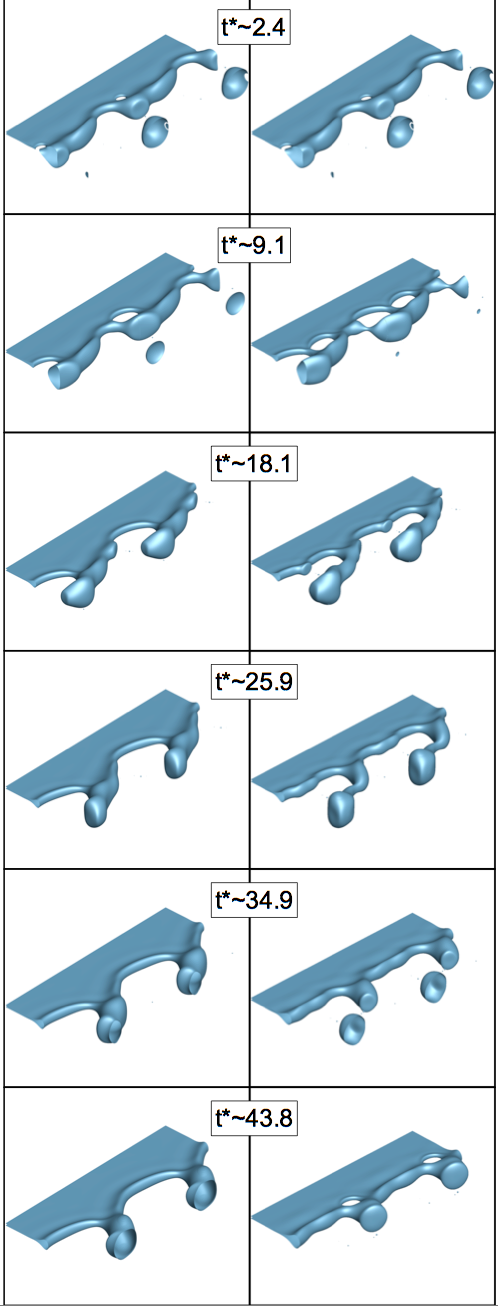}
\caption{The time-progression of moving domain results for retracting air films with $Oh=0.113$ and $Oh=0.072$ on the left and right respectively. Note that for visualization, the results are copied once in the z direction.} 
\label{3D_sim_oh113}
\end{figure}

For a water-air system, only one dimensionless number is required to parameterize the thin retracting gas film. We have chosen the Ohnesorge number ($Oh$) for this purpose. By changing the viscosity values of both phases while maintaining the viscosity ratio, we are able to alter $Oh$ and study its effect on the 3D dynamics of the film. So far we have presented 3D results for the nominal case, with $Oh=0.072$, which corresponds to a $h=2\mu m$ air film retracting in water. We use the same mesh and moving domain as the nominal case for all $Oh$ value simulations. The initial conditions and the boundary conditions are also the same as the nominal case. In particular, given the weak dependence of the retracting velocity on viscosity, the inflow velocity is the same for different $Oh$ value simulations. In Figure \ref{3D_sim_oh113} simulation results for $Oh=0.113$ are shown. This case corresponds to a $h=800nm$ air film in water and the initial conditions are the same as the nominal case moving domain simulation (Figure \ref{3D_sim_mm}). In Figure \ref{3D_sim_oh113}, a comparison of the evolution of the $Oh=0.113$ and $Oh=0.072$ films (on the left and right, respectively) is shown at the same non-dimensional times.
It seems that that viscous forces are more dominant in the $Oh=0.113$ simulations compared to the nominal case with $Oh=0.072$. These effects reduce the amplitude of the capillary waves (also observed in 2D, Figure \ref{2D_profiles_oh_113_32}) and make the interface smoother in general. Based on the mechanism explained for hole formation, holes are formed with less frequency in higher $Oh$ films, as can be seen in Figure \ref{3D_sim_oh113}. This may explain why these holes have not been explicitly observed in experiments where more viscous fluids are used to visualize microbubble shedding (Thoroddsen \textit{et al.} 2012). Moreover, the filaments are thicker for these films and seem to be more robust, causing delay of break-up and micro-bubble shedding. While the simulation is running, based on the volume in the film and the amount of time passed we have found a lower bound for the size of the micro-bubble resulting from the $Oh=0.113$ film:
\begin{equation}
{D}_{bubb,Oh=0.113}>18.4h
\label{Oh113_bubb_size}
\end{equation}
For a water-air system, the film thickness would be $800nm$, meaning that this micro-bubble has at least a diameter of $15\mu m$. We know from experiments that even thinner films result in micro-bubbles in the range of $10-100$ microns (Thoroddsen \textit{et al.}, 2012), therefore we expect ${D}_{bubb}/h$ to increase as $Oh$ is increased. Based on Equation \ref{Oh113_bubb_size}, our simulations are so far in line with this expectation. This simulation is also currently in progress, in addition to more expensive simulations at $Oh=0.226$ and $Oh=0.320$. By covering this wide range of $Oh$ numbers, we can deduct scaling laws and propose models for micro-bubble shedding from thin retracting gas films.


\section{Conclusions}
\label{SECconclusions}
In this work, we presented high resolution 2D and 3D numerical simulations of very long and thin retracting air films surrounded by water. For this pair of working fluids that are of interest for naval applications, the only parameter controlling the evolution of the film is its thickness. By fixing the density and viscosity ratios to match those of air and water, we studied the effect of the film thickness by varying the Ohnesorge ($Oh$) number of the film in the range of $0.072\le Oh\le 0.32$, corresponding to film thicknesses of $100nm\le h\le 2\mu m$ which are relevant based on drop-pool experimental measurements. From our 2D simulations we observed that bubbles do not shed from the gas films and 3D mechanisms must be responsible for micro-bubble shedding from thin gas films entrapped in drop-pool experiments. We also found that contrary to thin liquid films, the retraction speed of thin gas films does not reach a steady value and decreases with time as the rim becomes thicker. Based on the found retraction speed power law, we discovered that the 2D edge follows a similarity shape as it grows. Moreover, these 2D simulations enabled us to study the dependence of the rim shape and retraction speed on $Oh$ number. We found that in the range of $Oh$ numbers we studied, viscous effects do not cause significant variations in the two-dimensional dynamics. We then provided scaling arguments and referred to experiments to show that a capillary instability similar to the Rayleigh-Plateau instability can act upon the growing rim of these thin films if they have very high aspect ratios (long and thin films). Our 3D simulations confirmed this analysis, and showed that span-wise undulations appear on the rim of the film that cause puncture of the film. These rapidly growing holes distort the shape of the rim significantly and the slower retraction speed of larger parts of the rim eventually leads to their separation from the film, in the form of micro-bubbles. The process of micro-bubble shedding was thus found to be different from fragmentation of liquid films. These micro-bubbles were found to have diameters of around $15$ times the thickness of the film for $Oh=0.072$. Simulations at $Oh=0.113$ revealed that viscous effects delay hole formation by reducing capillary waves around the rim, resulting in larger bubbles with respect to the film thickness. In other words, our simulations indicate that micro-bubble size is a sub-linear function of film thickness. The 3D simulations are still in progress, from which we will be able to extract scaling laws and models that can be used for subgrid-scale modeling of micro-bubbles in practical two-phase simulations of flows around naval vessels.

\section{Acknowledgements}
\label{SECacknowledgements}
This work was supported by Office of Naval Research (Grant No. 119675). The authors gratefully acknowledge discussions and comments by Prof. Bud Homsy on an earlier version of our results.

\end{multicols*}

\begin{thebibliography}{00}
\bibitem{Agbaglah2013} Agbaglah, G., Josserand, C. and Zaleski, S., "Longitudinal instability of a liquid rim", \underline{Physics of Fluids}, Vol. 25, No. 2, p. 022103.
\bibitem{Aryafar2008} Aryafar, H. and Kavehpour, H.P., "Hydrodynamic instabilities of viscous coalescing droplets", \underline{Physical Review E}, Vol. 78, No. 3, 2008, p.037302.
\bibitem{Ashgriz1995} Ashgriz, N. and Mashayek, F., "Temporal analysis of capillary jet breakup", \underline{Journal of Fluid Mechanics}, 291, 1995, pp. 163-190.
\bibitem{Ashgriz2011} Ashgriz, N. ed., "Handbook of atomization and sprays: theory and applications", \underline{Springer Science \& Business Media}, 2011.
\bibitem{Brenner1999} Brenner, M.P. and Gueyffier, D., "On the bursting of viscous films", \underline{Physics of Fluids}, Vol. 11, No. 3, 1999, pp. 737-739.
\bibitem{Burton2005} Burton, J.C., Waldrep, R. and Taborek, P., "Scaling and instabilities in bubble pinch-off", Physical review letters, 94(18), 2005, p. 184502.
\bibitem{Chan2016} Chan, W.H.R., Urzay, J., Mani, A. and Moin, P., "On the development of a subgrid-scale model for the generation of micro-bubbles from impacting liquid surfaces", \underline{Center for Turbulence Research Annual Research Briefs}, 2016, pp. 149-162.
\bibitem{Chan2017} Chan, W.H.R., Urzay, J. and Moin, P., "Development of a collision detection algorithm for turbulent two-phase flows", \underline{Center for Turbulence Research Annual Research Briefs}, 2017, pp. 103-116.
\bibitem{Chandrasekhar} Chandrasekhar, S., "Hydrodynamic and hydromagnetic stability", \underline{Courier Corporation}, 2013.
\bibitem{Culick1960} Culick, F.E.C., "Comments on a ruptured soap film", \underline{Journal of applied physics}, Vol. 31, No. 6, 1960, pp. 1128-1129.
\bibitem{Deike2016} Deike, L., Melville, W.K. and Popinet, S., "Air entrainment and bubble statistics in breaking waves", \underline{Journal of Fluid Mechanics}, Vol. 801, 2016, pp. 91-129.
\bibitem{Esmaily2017} Esmaily, M., Jofre, L., Mani, A. and Iaccarino, G., "A scalable geometric multigrid solver for nonsymmetric elliptic systems with application to variable-density flows", \underline{Journal of Computational Physics}, 2017.
\bibitem{Esmailzadeh1986} Esmailizadeh, L. and Mesler, R., "Bubble entrainment with drops", \underline{Journal of colloid and interface science}, Vol. 110, No. 2, 1986, pp. 561-574.
\bibitem{Hendrix2016} Hendrix, M.H., Bouwhuis, W., van der Meer, D., Lohse, D. and Snoeijer, J.H., "Universal mechanism for air entrainment during liquid impact", \underline{Journal of fluid mechanics}, Vol. 789, 2016, pp.708-725.
\bibitem{Herrmann2010} Herrmann, M., "A parallel Eulerian interface tracking/Lagrangian point particle multi-scale coupling procedure", \underline{Journal of Computational Physics}, Vol. 229, No. 3, 2010, pp. 745-759.
\bibitem{Hicks2011} Hicks, P.D. and Purvis, R., "Air cushioning in droplet impacts with liquid layers and other droplets", \underline{Physics of Fluids}, Vol. 23, No. 6, 2011, p. 062104.
\bibitem{Keim2006}, N.C., Møller, P., Zhang, W.W. and Nagel, S.R., "Breakup of air bubbles in water: Memory and breakdown of cylindrical symmetry", \underline{Physical review letters}, 97(14), 2006, p. 144503.
\bibitem{Kim2011} Kim, D. and Moin, P., "Direct Numerical Simulation of Two-phase Flow with Application to Air Layer Drag Reduction", \underline{Doctoral dissertation, Stanford University}, 2011.
\bibitem{Krechetnikov2010} Krechetnikov, R., "Stability of liquid sheet edges", \underline{Physics of Fluids}, Vol. 22, No. 9, 2010, p. 092101.
\bibitem{Le1991} Le, H. and Moin, P., "An improvement of fractional step methods for the incompressible Navier-Stokes equations", \underline{Journal of computational physics}, Vol. 92, No. 2, 1991, pp. 369-379.
\bibitem{Mirjalili2014} Mirjalili, S. and Mani, A., "Thin sheet break-up in droplet-pool impact events", \underline{Bulletin of the American Physical Society}, Vol. 59, 2014.
\bibitem{Mirjalili2018a} Mirjalili, S., Ivey, C.B. and Mani, A., "On the boundedness properties of a conservative diffuse interface method for two-phase flows", \underline{arXiv preprint arXiv:1803.01262}, under review for publication in Journal of Computational Physics, 2018.
\bibitem{Mirjalili2018b} Mirjalili, S., Ivey, C.B. and Mani, A., "Comparison between the diffuse interface and volume of fluid methods for simulating two-phase flows", \underline{arXiv preprint arXiv:1803.07245}, 2018.
\bibitem{Mortazavi2016} Mortazavi, M., Le Chenadec, V., Moin, P. and Mani, A., "Direct numerical simulation of a turbulent hydraulic jump: turbulence statistics and air entrainment", \underline{Journal of Fluid Mechanics}, Vol. 797, 2016, pp. 60-94.
\bibitem{Oguz1989} Oguz, H.N. and Prosperetti, A., "Surface-tension effects in the contact of liquid surfaces", \underline{Journal of Fluid Mechanics}, Vol. 203, 1989, pp. 149-171.
\bibitem{Reed2002} Reed, A.M. and Milgram, J.H., "Ship wakes and their radar images", \underline{Annual Review of Fluid Mechanics}, Vol. 34, No. 1, 2002, pp. 469-502.
\bibitem{Reyssat2006} Reyssat, É. and Quéré, D., "Bursting of a fluid film in a viscous environment", \underline{EPL (Europhysics Letters)}, Vol. 76, No. 2, 2006, p.236.
\bibitem{Roisman2006} Roisman, I.V., Horvat, K. and Tropea, C., "Spray impact: rim transverse instability initiating fingering and splash, and description of a secondary spray", \underline{Physics of Fluids}, Vol. 18, No. 10, 2006, p. 102104.
\bibitem{Savva2009} Savva, N. and Bush, J.W., "Viscous sheet retraction", \underline{Journal of Fluid Mechanics}, Vol. 626, 2009, pp. 211-240.
\bibitem{Saylor2012} Saylor, J. and Bounds, G.D., "Experimental study of the role of the Weber and capillary numbers on Mesler entrainment", \underline{AIChE Journal}, Vol. 58, No. 12, 2012, pp. 3841-3851.
\bibitem{Sobyanin2015} Sob’yanin, D. N., "Theory of the antibubble collapse", \underline{Physical review letters}, 2015, 114(10), 104501.
\bibitem{Sunderhauf2002} S\"{u}nderhauf, G., Raszillier, H. and Durst, F., "The retraction of the edge of a planar liquid sheet", \underline{Physics of Fluids}, Vol. 14, No. 1, 2002, pp. 198-208.
\bibitem{Talor1959} Taylor, G., "The dynamics of thin sheets of fluid. II. Waves on fluid sheets", \underline{Proceedings of the Royal Society of London} , Series A, Mathematical and Physical Sciences, 1959, pp. 296-312.
\bibitem{Thoroddsen2003} Thoroddsen, S.T., Etoh, T.G. and Takehara, K., "Air entrapment under an impacting drop", \underline{Journal of Fluid Mechanics}, Vol. 478, 2003, pp. 125-134.
\bibitem{Thoroddsen2007} Thoroddsen, S.T., Etoh, T.G. and Takehara, K., "Experiments on bubble pinch-off", \underline{Physics of Fluids}, 19(4), 2007, p. 042101.
\bibitem{Thoroddsen2012} Thoroddsen, S.T., Thoraval, M.J., Takehara, K. and Etoh, T.G., "Micro-bubble morphologies following drop impacts onto a pool surface", \underline{Journal of Fluid Mechanics}, Vol. 708, 2012, pp. 469-479.
\bibitem{Tjahjadi1992} Tjahjadi, M., Stone, H.A. and Ottino, J.M., "Satellite and subsatellite formation in capillary breakup", \underline{Journal of Fluid Mechanics}, 243, 1992, pp. 297-317.
\bibitem{Tran2013} Tran, T., de Maleprade, H., Sun, C. and Lohse, D., "Air entrainment during impact of droplets on liquid surfaces", \underline{Journal of Fluid Mechanics}, Vol. 726, 2013.
\bibitem{Trevorrow1994} Trevorrow, M. V., Vagle, S. and Farmer, D. M., "Acoustical measurements of microbubbles within ship wakes", \underline{The Journal of the Acoustical Society of America}, Vol. 95, No. 4, 1994, pp. 1922-1930.
\bibitem{Wang2016} Wang, Z., Yang, J. and Stern, F., "High-fidelity simulations of bubble, droplet and spray formation in breaking waves", \underline{Journal of Fluid Mechanics}, Vol. 792, 2016, pp. 307-327.
\bibitem{Zaleski2017} Zaleski, S., Ling, Y., Fuster, D., and Tryggvason, G., "Atomisation and droplet formation mechanisms in a model two-phase mixing layer", \underline{Bulletin of the American Physical Society}, 2017, 62.
\end{thebibliography}
\end{document}